\documentclass[10pt]{article}
\usepackage[utf8]{inputenc}

\usepackage{amsmath}
\usepackage{amsfonts}
\usepackage{amssymb}
\usepackage{makeidx}
\usepackage{graphicx}
\usepackage{caption}
\usepackage{subcaption}
\usepackage{listings}
\usepackage{apacite}
\usepackage{natbib}
\usepackage{indentfirst}
\usepackage{bbm}
\usepackage{xcolor}
\usepackage{siunitx}
\usepackage{multirow}
\usepackage{booktabs}
\usepackage{mathtools}
\usepackage{enumitem}
\usepackage{setspace}
 
\usepackage[margin=1in]{geometry}
\usepackage{authblk}
\usepackage[pdfborder={0 0 0}]{hyperref}

\title{Option Pricing Incorporating Factor Dynamics in Complete Markets}

\author[a]{Yuan Hu}
\author[b]{Abootaleb Shirvani}
\author[c]{W. Brent Lindquist}
\author[d]{Frank J. Fabozzi}
\author[e]{Svetlozar T. Rachev}

\affil[a]{\small Texas Tech University\\
\url{yuan.hu@ttu.edu}}
\affil[b]{\small Texas Tech University\\
\url{abootaleb.shirvani@ttu.edu}}
\affil[c]{\small Texas Tech University\\
\url{brent.lindquist@ttu.edu}}
\affil[d]{\small EDHEC Business School\\
\url{fabozzi@edhec.edu}}
\affil[e]{\small Texas Tech University\\
\url{zari.rachev@ttu.edu}}
\begin{document}

\thispagestyle{plain}
\begin{spacing}{1.0}
\maketitle
\noindent {\textbf {Abstract}}\ \ 
Using the Donsker-Prokhorov invariance principle we extend the Kim-Stoyanov-Rachev-Fabozzi option pricing model to allow for
variably-spaced trading instances, an important consideration for short-sellers of options.
Applying the Cherny-Shiryaev-Yor invariance principles, we formulate a new binomial path-dependent pricing model for discrete- and
continuous-time complete markets where the stock price dynamics depends on the log-return dynamics of a market influencing factor.
In the discrete case, we extend the results of this new approach to a financial market with informed traders employing a statistical
arbitrage strategy involving trading of forward contracts.
Our findings are illustrated with numerical examples employing US financial market data.
Our work provides further support for the conclusion that any option pricing model must preserve valuable information on the
instantaneous mean log-return, the probability of the stock's upturn movement (per trading interval),
and other market microstructure features.
\end{spacing}

\section{Introduction}
\label{sec1}
\noindent The Donsker-Prokhorov invariance principle (DPIP), also known as the Functional Limit Theorem,
is a fundamental result in the theory of stochastic processes and a limit theorem for sequences of random variables.\footnote{
	See \cite{Donsker1951}; \cite{Prokhorov1956}; \cite{Major1978}; \cite{Billingsley1999}, Section 14; \cite{Gikhman1969},
	Chapter IX; \cite{Skorokhod2005}, Section 5.3.3; \cite{Davydov2008}; and \cite{Fischer2011}, Section 7.3.}
\cite{Cox1979} were the first to use DPIP  in their seminal Cox-Ross-Rubinstein (CRR)-binomial option pricing model.\footnote{
	\cite{Cox1979} used the central limit theorem for triangular series, which can be viewed as a special case of DPIP.}
Although there are several extensions of the CRR-model\footnote{
	See \cite{Jarrow1983}; \cite{Hull2018}; \cite{Boyle1986}; \cite{Boyle1988}; \cite{Madan1989}; \cite{Rubinstein2000};
	\cite{Jabbour2005}; \cite{Whaley2006}; \cite{Florescu2008}; \cite{Yuen2013}; \cite{Sierag2018}.},
rigorous proofs of the corresponding limit results leading to continuous-time option pricing formula are often not provided.\footnote{
	On rare occasions, the weak convergence of the c\`{a}dl\`{a}g price process generated by the binomial (trinomial, or multinomial)
	pricing tree is shown.}
In this paper, we provide the proofs for various extensions of DPIP to obtain a variety of new binomial option pricing models.
A second, but far more disturbing, issue observed  in the literature on binomial option pricing is the often-seen statement
that binomial option pricing does not depend on the underlying stock mean log-return and the probability for upward movement in
the binomial model.
This leads to the {\it binomial discontinuity option price puzzle}, based on the claim that regardless of how close the natural
(historical) probability $p_{\Delta t}\in(0,1)$ of a stock's upturn movement (in a given trading period $\Delta t$) is to $1$ or $0$,
the binomial option price stays unchanged, but when $p_{\Delta t} = 1$ (or  $p_{\Delta t} = 0$) the option price jumps to the
price of a risk-free asset.
This erroneous conclusion results from the third step in the following 3-step sequence of arguments in binomial option pricing.
\begin{itemize}[leftmargin=35pt]
	\item[Step 1:] Introduce a continuous-time arbitrage-free model for the underlying stock price, for example,
	a geometric Brownian motion with instantaneous mean log-return $\mu>r$ and volatility  $\sigma >0$, where $r$ is the risk-free rate.
	\item[Step 2:] From the arbitrage-free asset pricing model\footnote{
		This step is based on the Fundamental Theorem of Asset Pricing,	see \cite{Delbaen1998}, \cite{Duffie2001}, Chapter 6.},
	obtain the continuous-time risk-neutral option price dynamics.
	In the case of geometric Brownian motion in the first step, these dynamics depend only on $r$, $\sigma$,
	and the option's contract-specifications. 
	\item[Step 3:] Construct a binomial tree on the risk-neutral world, ensuring convergence\footnote{
		The convergence is either (a) in terms of one-dimensional distributions of the process generated by the risk-neutral
	pricing tree, or (b) in terms of multivariate distributions, or (c) in Skorokhod $\mathcal{D}[0,T]$-topology.}
	of the pricing tree to the limiting, risk-neutral, continuous-time price process.
\end{itemize}
Obviously, the third step of binomial option pricing is misguided. In continuous-time option valuation, the self-financing portfolio,
which replicates the option value, can be updated continuously in time without any transaction cost, which is an absurdity in any real trading.
As a result, regardless of whether  $\mu$ goes to $\pm\infty$, the option price stays unchanged.
Due to Step 3, the binomial option pricing formula loses valuable information about the mean log-return $\mu$
and the probability $p_{\Delta t}$.
As shown in Kim {\it{et al.}} (2016, 2019) and \cite{Hu2020}, preserving $\mu$ and $p_{\Delta t}$ can be achieved in complete market
binomial models by (a) determining the delta-position in the underlying stocks using the arbitrage-free argument,
and then (b) passing to risk-neutral option valuation without using any continuous-time option model. 

All risk-neutral tri- and multinomial option pricing models, starting directly in the risk-neutral world,
approximate the dynamics of the continuous-time risk-neutral pricing process.
The tri- and multinomial approaches leave unanswered the question of which discrete arbitrage-free pricing model in the natural world
leads to the corresponding discrete risk-neutral pricing model.
One way to resolve this issue is to replace risk-neutral hedging with mean-variance hedging.\footnote{
	See \cite{Duffie1991}; \cite{Yamada2004}.}
This approach is generally used when the market for the underlying is incomplete.
In this paper, we only deal with risk-neutral hedging in our general binomial pricing models.

To illustrate our approach in resolving the issue of a binomial option pricing formula being independent\footnote{
See, for example, \cite{Hull2018}, Chapters 13.} of $\mu$ and $p_{\Delta t}\in(0,1)$, we consider the basic one-period option pricing model.
The stock price at the terminal time $\Delta t$ is given by\footnote{
	The same argument can be extended for any  $\Delta t >0$; see \cite{Rendleman1979}, and \cite{Hu2020} for various extensions.
	We denote “with probability” as “w.p.” for brevity.}
\begin{equation*}
S_{\Delta t} = \begin{cases} 
      S_0 u & \textup{w.p.} \; p_{\Delta t}\\
      S_0 d & \textup{w.p.} \; 1-p_{\Delta t} 
   \end{cases},\;p_{\Delta t} \in (0,1),\; o(\Delta t) = 0,
\end{equation*}
with the log-return time series $R_{\Delta t} = \ln\left(S_{\Delta t}/S_0\right)$. Choose parameters $u$ and $d$, so that $\mathbb{E}\left(R_{\Delta t}\right) = \mu \Delta t$ and $\textup{Var}\left(R_{\Delta t}\right) = \sigma^2\Delta t,\; \sigma>0$. This implies
\begin{equation*}
	u = 1+\mu \Delta t+ \sigma\sqrt{\frac{1-p_{\Delta t}}{p_{\Delta t}}}\sqrt{\Delta t},\;\;\;
	d = 1+\mu \Delta t- \sigma\sqrt{\frac{p_{\Delta t}}{1-p_{\Delta t}}}\sqrt{\Delta t}.
\end{equation*}
Then, the arbitrage-free argument leads to the one-period option price
\begin{equation*}
	f_0 = e^{-r\Delta t}\left(q_{\Delta t}f_{\Delta t}^{(u)}+(1-q_{\Delta t})f_{\Delta t}^{(d)}\right),
\end{equation*}
where $q_{\Delta t} = p_{\Delta t} - \theta \sqrt{p_{\Delta t}(1-p_{\Delta t})\Delta t}$ and $\theta = (\mu-r)/\sigma$ is the market price of risk.\footnote{See \cite{Kim2016} and \cite{Kim2019} for the multiperiod extension.}
The comment in \cite{Hull2018} “The option pricing formula in equation (13.2) does not involve the probabilities of the stock price moving up or down.” is erroneous.
As a matter of fact, if $p_{\Delta t}\uparrow 1$, then $q_{\Delta t}\uparrow 1$, and if $p_{\Delta t} \downarrow
0$, then $q_{\Delta t} \downarrow 0$, and this observation resolves the binomial discontinuity option price puzzle.
Furthermore, the risk-neutral probability $q_{\Delta t}$ does depend on $\mu$.
Addressing the misconception about binomial option pricing being independent of $\mu$ and $p_{\Delta t}$ is the main motivation for the results in
Sections \ref{sec2} and \ref{sec3} in this paper.
In these two sections, general binomial option pricing formulas are derived and the corresponding continuous-time limits are shown by applying DPIP.
Motivated by the Ross Recovery Theorem\footnote{See \cite{Ross2015}, \cite{Audrino2019}, and \cite{Jackwearth2020}.},
the {\it implied $p_{\Delta t}$-surface} is introduced and estimated on real financial data.
The main conclusion derived from Sections \ref{sec2} and \ref{sec3} is that the information on $p_{\Delta t}$ and $\mu$ should be used in binomial
option pricing.
Estimating $p_{\Delta t}$ and $\mu$ from historical data and using those estimates in binomial option pricing could deliver more flexible and realistic
option pricing models.\footnote{See \cite{Johnson1997} and \cite{Yamada2004} for similar considerations.}

Next, we consider possible applications of non-standard invariance principles\footnote{For example, \cite{Billingsley1962}; \cite{Silvestrov2004}, Chapter 4; \cite{Gut2009}, Chapter 5; \cite{Tanaka2017}, Chapter 2; \cite{Cherny2003}; and \cite{Crimaldi2019}.} to  option pricing theory. We are convinced that non-standard invariance principles should serve as a great source for introducing new types of valuable discrete-time option pricing models. These new discrete-time option pricing models, together with the corresponding limiting continuous-time option pricing models, could exhibit features already observed in empirical studies on option pricing but not presented in the current theoretical option pricing models. In Sections \ref{sec4}, \ref{sec5} and \ref{sec6}, we illustrate the usefulness (to the theory of option pricing) of one non-standard invariance principle, the {\it Cherny-Shiryaev-Yor Invariance Principle} (CSYIP) by \cite{Cherny2003}. In Section \ref{sec4}, applying an extension of the CSYIP, we derive a new binomial stock pricing model where the underlying stock price depends on the log-return trajectory of another stock, or stock-index, or any observable risk factor influencing the underlying stock dynamics. We provide a numerical illustration of the model. In Section \ref{sec5}, we derive a new option price formula based on the stock binomial price model introduced in Section \ref{sec4}. We estimate the volatility surface based on this new option price model. In Section \ref{sec6} we extend the results in Section \ref{sec5} to markets with informed (and misinformed) traders. Here we follow the general framework of a financial market with informed, misinformed, and noisy traders introduced and studied in \cite{Hu2020}. Using the call option where the underlying stock is Microsoft (MSFT)\footnote{The options data are from CBOE options on MSFT, see {\it www.cboe.com/MSFT}.}, we estimate the {\it implied information rate} of MSFT call option traders. Our conclusions are summarized in Section \ref{sec7}.

\section{Donsker-Prokhorov Invariance Principle and Binomial Option Pricing}
\label{sec2}

\noindent
The classic CRR-binomial option pricing and Jarrow-Rudd (JR)-binomial
models\footnote{See \cite{Cox1979} for the CRR-binomial option pricing model;
\cite{Jarrow1983} and \cite{Hull2018} for the JR-binomial option pricing model.}
do not include the probability $p_{\Delta t}$ for a stock's upturn as a parameter.
The KSRF enhanced binomial model \citep{Kim2016}, extended CRR- and JR-binomial models to include $p_{\Delta t}$ as a model parameter.
In this section, we apply DPIP to further extend the KSRF-model to allow for variably-spaced trading instances, which, for example, is of critical importance to the short seller of an option.

Consider the classic Black-Sholes-Merton market model (\citealp{Black1973}; \citealp{Merton1973}).
Assume the dynamics of the risky asset\footnote{The risky asset is the stock and will be denoted by $\mathcal{S}$.}
follow geometric Brownian motion,
\begin{equation}
	S_t = S_0 \exp \left\{(\mu-\frac{\sigma^2}{2})t+\sigma B_t\right\},\;S_0 >0,\; \mu>0,\;\sigma>0, \;t \in [0,T].
	\label{eq1_gbm}
\end{equation}
In (\ref{eq1_gbm}), the price process $S_t$ is defined on a stochastic basis $\left(\Omega,\mathbb{F} = \left\{\mathcal{F}_t = \sigma(B_u)\right\}_{u \leq t,0 \leq t\leq T},\mathbb{P}\right)$ generated by the standard Brownian motion (BM): $\mathbb{B}^{[0,T]} = \left\{B_t\right\}_{0 \leq t\leq T}$. The dynamics of the risk-free asset\footnote{The risk-free asset will be a bond and will be denoted by $\mathcal{B}$.} with the risk-free rate $r$  is given by
\begin{equation}
	\beta_t = \beta_0 e^{rt},\; \beta_0>0,\;\mu>r>0,\; t \in [0,T].
	\label{eq2_bond}
\end{equation}
Let $f_t = f(S_t,t),\; t\in[0,T]$ be the price dynamics of a European Contingent Claim (ECC)\footnote{The option or derivative will be denoted by $\mathcal{C}$.} with terminal time $T>0$ and final payoff $f_T = G(S_T)$.
We construct a general binomial pricing model applying DPIP.

\subsection{Donsker-Prokhorov invariance principle}
\label{sec21}
\noindent The following version of DPIP is based on the \cite{Davydov2008} “non-classical” treatment of the DPIP.

\noindent({\bf DPIP}) {\it Consider two sets of triangular series of random variables
$\xi_{n,1},\ldots,\xi_{n,k_n},\;\eta_{n,1},\ldots,\eta_{n,k_n},\; n\in \mathcal{N} = \{1,2,\ldots\}$, satisfying the following conditions:
\begin{itemize}
	\item[(i)] for every $n\in \mathcal{N},\;\xi_{n,1},\ldots,\xi_{n,k_n},$ is a sequence of independent random variables,
and $\eta_{n,1},\ldots,\eta_{n,k_n}$ is a sequence of independent normal random variables;
	\item[(ii)] $\mathbb{E}(\xi_{n,k}) = \mathbb{E}(\eta_{n,k}) = a_{n,k},\quad
Var(\xi_{n,k}) =Var(\eta_{n,k}) = \sigma^2_{n,k}, \quad k = 1,\ldots,k_n$.
Let
	\begin{align*}
	\mathbb{X}_n^{[0,T]} &= \left\{ X_n(t) = X_{n,k},\; t\in [t_{n,k},t_{n,k+1}),\;k = 0,\ldots,k_{n-1},
\quad X_n(T) = X_{n,k_n},\;X_n(0) = X_{n,0}\right\},
	\\
	\mathbb{Y}_n^{[0,T]} &= \left\{ Y_n(t) = Y_{n,k},\; t\in [t_{n,k},t_{n,k+1}),\;k = 0,\ldots,k_{n-1},
\quad Y_n(T) = Y_{n,k_n},\;Y_n(0) = Y_{n,0}\right\},
\end{align*}
where:
$X_{n,k} = \sqrt{\frac{T}{c_n}}\sum^{k}_{i=1}(\xi_{n,i}-a_{n,i}),\;X_{n,0} =0$;
$Y_{n,k} = \sqrt{\frac{T}{c_n}}\sum^{k}_{i=1}(\eta_{n,i}-a_{n,i}),\;Y_{n,0} =0$;\hfil\break
$c_n = \sum^{k_n}_{i=1}\sigma^2_{n,i}$;
and $t_{n,k} = \frac{T}{c_n}\sum^{k}_{i=1}\sigma^2_{n,i},\;t_{n,0} = 0$, for $k = 1,\ldots,k_n$.
\end{itemize}}

\noindent
({\bf DPIP1}) {\it If
\begin{equation}
	\lim_{n \uparrow \infty}\sum^{k_n}_{i=1}\int_{\{x:|x|>\epsilon\}}\left|x\right|
\left|\mathbb{P}\left(\xi_{n,i}-a_{n,i}\leq x\sqrt{\frac{c_n}{T}}\right)
- \mathbb{P}\left(\eta_{n,i}-a_{n,i}\leq x\sqrt{\frac{c_n}{T}}\right)\right|dx = 0,\; \forall \epsilon>0,
	\label{eq3_dpip1}
\end{equation}
then, $\lim_{n \uparrow \infty}FM(\mathbb{X}_n^{[0,T]},\mathbb{Y}_n^{[0,T]}) = 0$,
where $FM(\mathbb{X},\mathbb{Y}) = \underset { \mathbbm{f} } { \sup } \left\{\left| \mathbb{E}\left(\mathbbm{f}(\mathbb{X})-\mathbbm{f}(\mathbb{Y})\right)\right|:\mathbbm{f} \in \mathcal{FM}  \right\}$\footnote{
See \cite{Rachev1985}, $\mathcal{FM} = \left\{\mathbbm{f}:\mathcal{D}[0,T]\rightarrow R:\; \Vert \mathbbm{f} \Vert _{FM} =\sup_{\mathbbm{a},\mathbbm{b}\in \mathcal {D}[0,T],\mathbbm{a}\neq\mathbbm{b}}\frac{|\mathbbm{f}(\mathbbm{a})-\mathbbm{f}(\mathbbm{b}) |}{d^{(0)}(\mathbbm{a},\mathbbm{b}) \max \left(1,d^{(0)}(\mathbbm{a},\mathbbm{o}),d^{(0)}(\mathbbm{b},\mathbbm{o})\right)}\leq 1\right\}$,
where $\mathbbm{o}(t) = 0,\;t \in[0,T]$.
The Skorokhod space $\mathcal{D}[0,T]$ is the space of all functions $a:[0,T]\rightarrow R$ that are right continuous with left-hand limits (c\`{a}dl\`{a}g functions), and $d^{(0)}(\mathbbm{a},\mathbbm{b}) = \inf_{\lambda \in \Lambda} \max \left\{ \sup_{0\leq s<t\leq T} |\ln \frac{\lambda(t)-\lambda(s)}{t-s}|,\sup_{0\leq t\leq T}|\mathbbm{a}(t)-\mathbbm{b}\left(\lambda(t)\right)| \right\}$,
which is the Skorokhod-Billigsley metric in $\mathcal{D}[0,T]$ with $\Lambda = \left\{\textup{all strictly increasing continuous functions } \lambda:[0,T]\rightarrow [0,T],\; \lambda(0) = 0,\;\lambda(T) = T \right\}$ being the Skorokhod-Billingsley metric in $\mathcal{D}[0,T]$; see Section 12 of \cite{Billingsley1999}; \cite{Skorokhod1956}.}
is the Fortet-Mourier metric\footnote{See \cite{Fortet1953}; \cite{Dudley2018}.} in the Polish (complete separable metric) space of random elements
with values in the Skorokhod space $(\mathcal{D}[0,T],d^{(0)})$}.

\noindent
({\bf DPIP2}) {\it If
\begin{align*}
	\mathbbm{m}_n = \max_{k=1,\ldots,k_n}\left\{\Delta t_{n,k} = t_{n,k}-t_{n,k-1} = T\frac{\sigma^2_{n,k}}{c_n}  \right\} \rightarrow 0, \; as\; n \uparrow \infty, 
\end{align*}
then, $\lim_{n \uparrow \infty}FM(\mathbb{Y}_n^{[0,T]},\mathbb{B}^{[0,T]}) = 0$. Furthermore, condition (\ref{eq3_dpip1}) matches Lindeberg's condition:
\begin{equation}
	\lim_{n\uparrow\infty}\frac{1}{c_n}\sum_{i=1}^n\mathbb{E}(\xi_{n,i}-a_{n,i})^2I_{\left\{\left|\xi_{n,i}-a_{n,i}\right|>\epsilon\sqrt{c_n}\right\}} = 0,\;\forall \epsilon>0,
	\label{eq4_lindeberg}
\end{equation}
which leads to $\lim_{n\uparrow\infty}FM(\mathbb{X}_n^{[0,T]},\mathbb{B}^{[0,T]}) = 0$.
}

The $FM$-convergence of random elements with values in $(\mathcal{D}[0,T],d^{(0)})$ is characterized by the following proposition. 

\noindent
({\bf Convergence in the metric \textit{FM}}) {\it Let $\mathbb{X}_n,\;n\in \mathcal{N}$ and $\mathbb{Y}$ be random elements with values in $(\mathcal{D}[0,T],d^{(0)})$ with $\mathbb{E}\left(d^{(0)}(\mathbb{X}_n,\mathbbm{o})^2\right)< \infty,$ and $\mathbb{E}\left(d^{(0)}(\mathbb{Y},\mathbbm{o})^2\right)< \infty$. Then the following statements are equivalent:
\begin{itemize}
\item[(i)] $\lim_{n\uparrow\infty}FM(\mathbb{X}_n,\mathbb{Y}) = 0$;
\item[(ii)] $\mathbb{X}_n$ weakly converges \citep{Billingsley1999} to $\mathbb{Y}$ as $n \rightarrow \infty$, and $\lim_{n\uparrow\infty}\mathbb{E}(d^{(0)}(\mathbb{X}_n,\mathbbm{a})^q) = \mathbb{E}(d^{(0)}(\mathbb{Y},\mathbbm{a})^q)$, for every $q \in (0,2]$ and $\mathbbm{a}\in \mathcal{D}[0,T]$;
\item[(iii)] There exists\footnote{See \cite{Skorokhod1956} and \cite{Dudley1968}.} a probability space $(\Omega^*,\mathcal{F}^*,\mathbb{P}^*)$ with random elements $\mathbb{X}_n^*,\;n\in \mathcal{N}$ and $\mathbb{Y}^*$ with values in $(\mathcal{D}[0,T],d^{(0)})$ such that (a) the probability law of $\mathbb{X}_n^*$ coincides with the probability law of $\mathbb{X}_n$ $(\mathbb{P}^*_{\mathbb{X}_n^*} = \mathbb{P}_{\mathbb{X}_n})$, for all $n \in \mathcal{N}$; (b) $\mathbb{P}^*_{\mathbb{Y}^*} = \mathbb{P}_{\mathbb{Y}}$; (c) $\mathbb{P}^*\left(\lim_{n\rightarrow \infty}\mathbb{X}_n^* =\mathbb{Y}^* \right) = 1$; and (d) $\lim_{n\uparrow\infty}\mathbb{E}^{\mathbb{P}^*}(d^{(0)}(\mathbb{X}_n,\mathbbm{a})^q) = \mathbb{E}^{\mathbb{P}^*}(d^{(0)}(\mathbb{Y},\mathbbm{a})^q)$, for every $q \in (0,2]$ and $\mathbbm{a}\in \mathcal{D}[0,T]$.
\end{itemize}
}

\noindent
({\bf Dual representation for \textit{FM}}) {\it  The following  Kantorovich mass-transshipment duality representation holds:
\begin{equation*}
	FM(\mathbb{X},\mathbb{Y}) = \inf \left\{\int_{\mathcal{D}[0,T]\times \mathcal{D}[0,T]}d^{(0)}(\mathbbm{a},\mathbbm{b}) \max \left(1,d^{(0)}(\mathbbm{a},\mathbbm{o}),d^{(0)}(\mathbbm{b},\mathbbm{o})\right)Q(d\mathbbm{a}\times d\mathbbm{b}),Q\in \mathbb{Q}(\mathbb{P}_{\mathbb{X}},\mathbb{P}_{\mathbb{Y}})  \right\},
\end{equation*}
where $\mathbb{Q}(\mathbb{P}_{\mathbb{X}},\mathbb{P}_{\mathbb{Y}})$ is the space of all positive Borel measures on $\mathcal{D}[0,T]\times \mathcal{D}[0,T]$ satisfying the marginal condition,
\begin{equation*}
	\mathbb{Q}\left(\mathbb{A}\times\mathcal{D}[0,T]) - \mathbb{Q}(\mathcal{D}[0,T]\times \mathbb{A}\right) = \mathbb{P}_{\mathbb{X}}(\mathbb{A})-\mathbb{P}_{\mathbb{Y}}(\mathbb{A})
\end{equation*}
for any Borel set $\mathbb{A}$ in $(\mathcal{D}[0,T],d^{(0)})$.}

According to DPIP1, in general the weak limits for $\mathbb{X}_n^{[0,T]}$  consist of connected segments of BM.\footnote{See \cite{Davydov2008}.} As those are not necessarily semimartingales, the study of binomial option pricing based on DPIP1, while of interest\footnote{Applying DPIP1 to binomial option pricing will allow the trading of the replication portfolios to occur in time-instances with general time-varying trading frequency $F(t,\Delta t),\;t\in[0,T]$, which is an important issue in market microstructure. To go beyond the dynamic asset pricing model of \cite{Delbaen1998} will require approaches based upon market microsctructure (See, e.g. \cite{Ohara1998}; and \cite{Hasbrouk2007}).
We believe that DPIP1 will provide a bridge between dynamic asset pricing and market microstructure. Applying DPIP2 restricts the trading frequency to  $F(t,\Delta t)=\Delta t + o(\Delta t),\;t\in[0,T]$.}, requires significantly different dynamic asset pricing methods and will not be discussed in the current paper.\footnote{\cite{Coviello2011} discussed dynamic asset pricing models based on so-called A-semimartingales. A-semimartingales can appear as limits in the DPIP1. This paper falls within the framework of reconciling dynamic asset pricing theory with market microstructure.} We next apply DPIP2 to a general binomial pricing model.

\subsection{General binomial pricing tree and their limits via DPIP2}
\label{sec22}
\noindent Consider a triangular array of binary random variables $\varsigma_{n,1},\varsigma_{n,2},\ldots,\varsigma_{n,k_n},\; n\in \mathcal{N}$,
satisfying the following conditions:
(a) for every $n\in \mathcal{N}$, $\varsigma_{n,1},\varsigma_{n,2},\ldots,\varsigma_{n,k_n}$ is a sequence of independent random variables;
and (b) $\mathbb{P}(\varsigma_{n,k} =1) =1-\mathbb{P}(\varsigma_{n,k} =-1) = p_{n,k}\in (0,1),\; k =1,\ldots,k_n,\; n\in \mathcal{N}$.
Consider the discrete filtration $\mathbb{F}^{(n)} = \left\{\mathcal{F}^{(n,0)} = \{\varnothing,\Omega\},\mathcal{F}^{(n,k)} = \sigma(\varsigma_{n,1},\ldots,\varsigma_{n,k}),k=1,\ldots,k_n\right\}$.
Define the following general binomial pricing tree\footnote{See \cite{Kim2016} and \cite{Kim2019}.}.

\noindent
({\bf General Binomial Pricing Tree}) For $k=1,\ldots,k_n,\;n\in \mathcal{N}$,
\begin{itemize}
	\item[(i)] let $t_{n,k},\;0=t_{n,0}< \cdots <t_{n,k}< \cdots <t_{n,k_n}=T$, be the times at which an option trader,
who holds a short position in the $\mathcal{C}$-contract, effects trades; and
	\item[(ii)] define the $n^{th}$-pricing tree, $n\in \mathcal{N}$, which
(a) is $\mathbb{F}^{(n)}$-adapted,
(b) is determined by the nodes $S_{t_{n,k}}^{(n)},\;k=1,\ldots,k_n,\;S_{t_{n,0}}^{(n)} = S_0$, and
(c) has the price dynamics,
\begin{equation}
	S^{(n)}_{t_{n,k}} =
	\begin{cases}
	S^{(n,u)}_{t_{n,k}}&=S^{(n)}_{t_{n,k-1}}\exp\left\{\left(\mu- \frac{1-p_{n,k}}{p_{n,k}}\frac{\sigma^2}{2}\right)\Delta t_{n,k}+\sigma\sqrt{\frac{1-p_{n,k}}{p_{n,k}}}\sqrt{\Delta t_{n,k}} \right\},\;\textup{if}\;\varsigma_{n,k} = 1,
	\\
	S^{(n,d)}_{t_{n,k}}&=S^{(n)}_{t_{n,k-1}}\exp\left\{\left(\mu- \frac{p_{n,k}}{1-p_{n,k}}\frac{\sigma^2}{2}\right)\Delta t_{n,k}-\sigma\sqrt{\frac{p_{n,k}}{1-p_{n,k}}}\sqrt{\Delta t_{n,k}} \right\},\;\textup{if}\;\varsigma_{n,k} = -1,
	\end{cases}
	\label{eq5_b_tree}
\end{equation}
where $\Delta t_{n,k} = t_{n,k} - t_{n,k-1},\;k=1,\ldots,k_n$. In (\ref{eq5_b_tree}), $p_{n,k}\in(0,1),\;k=1,\ldots,k_n,\;n\in\mathcal{N}$
 is the probability for upward movement of the stock price in $\Delta t_{n,k}$.
\end{itemize}
\noindent
Let: $\xi_{n,k} = R_{t_{n,k}}^{(n)}\vcentcolon = \ln\left(S^{(n)}_{t_{n,k}}/S^{(n)}_{t_{n,k-1}}\right),\;k =1,\ldots,k_n,\;n\in\mathcal{N}$;
$a_{n,k} = \mathbb{E}\left(\xi_{n,k}\right) = \left(\mu-\frac{\sigma^2}{2}\right)\Delta t_{n,k}$;
and $\sigma_{n,k}^2 = \textup{Var}\left(\xi_{n,k}\right) = \sigma^2 \Delta t_{n,k}$.
Thus, $c_n = \sum_{i=1}^{k_n}\sigma^2_{n,i} = \sigma^2 T$
and $t_{n,k} = \frac{1}{c_n}\sum_{i=1}^{k}\sigma_{n,i}^2,\;k =1,\ldots,k_n,\;t_{n,0}=0$.
If we assume that $\max_{k=1,\ldots,k_n} \Delta t_{n,k} = O(T/n)$
and $\max_{k = 0,\ldots,k_n-1} | p_{n,k} - p_{n,k+1}| = O(T/n)$, then the tree is recombining.

Consider the discrete log-return process,
\begin{equation*}
	\mathbb{R}_n^{[0,T]} = \left\{\mathbb{R}_t^{(n)} = \left\{R^{(n)}_{t_{n,k}},\;t\in[t_{n,k},t_{n,k+1}),k = 0,\ldots,k_n-1\right\},\mathbb{R}_0^{(n)} = R^{(n)}_{t_{n,0}} = 0,\; R_T^{(n)} = R_{t_{n,k_n}}^{(n)}\right\}.
\end{equation*}
In order to satisfy Lindeberg’s  condition (\ref{eq4_lindeberg}), we assume that $p_{n,k} = p^{(0)}+p^{(1)}\sqrt{\Delta t_{n,k}}+p^{(2)}\Delta t_{n,k} \in (0,1),\; k=1,\ldots,k_n,\; n\in \mathcal{N}$, for some $p^{(0)}\in (0,1),\; p^{(1)}\in R,$ and $p^{(2)}\in R$. Then, $\forall \epsilon>0$,
\begin{equation*}
	\lim_{n\uparrow\infty}\frac{1}{c_n}\sum_{i=1}^{k_n}\mathbb{E}(R_{t_{n,i}}^{(n)}-a_{n,i})^2I_{\left\{\left|R_{t_{n,i}}^{(n)}-a_{n,i}\right|>\epsilon\sqrt{c_n}\right\}} =\lim_{n\uparrow\infty}I_{\left\{\max\left(\frac{1-p^{(0)}}{p^{(0)}},\frac{p^{(0)}}{1-p^{(0)}}\right)>\epsilon^2\right\}}\sigma^2T= 0\;.
\end{equation*}
\noindent
Let $\; X_{n,k} = \frac{1}{\sigma}\left[\sum_{i=1}^k R_{t_{n,i}}^{(n)} - \left(\mu-\frac{\sigma^2}{2}\right)t_{n,k}\right]$ for $t\in[t_{n,k},t_{n,k+1}),\; k=1,\ldots,k_n,$
and $X_n(T) = X_{n,k_n}$.
Consider  the log-return process in continuous-time 
\begin{equation*}
	\mathbb{R}_n^{[0,T]} = \left\{\mathbbm{R}_t= \ln \left(\frac{S_t}{S_0}\right) = \left(\mu-\frac{\sigma^2}{2}\right)t+\sigma B_t,\; t\in[0,T]\right\}.
\end{equation*}
We set $\eta_{n,i} = \mathbbm{R}_{t_{n,i}}-\mathbbm{R}_{t_{n,i-1}},\;i=1,\ldots,k_n$.
Applying DPIP2, we have $\lim_{n\uparrow\infty}FM(\mathbb{X}_n^{[0,T]},\mathbb{B}^{[0,T]}) = 0$. Furthermore, there exists a space $(\Omega^*,\mathcal{F}^*,\mathbb{P}^*)$ with random elements $\mathbb{X}_n^{[0,T]*},\;n\in \mathcal{N}$ and $\mathbb{Y}^{[0,T]*}$ with values in $(\mathcal{D}[0,T],d^{(0)})$ such that $\mathbb{P}^*_{\mathbb{X}_n^{[0,T]*}} = \mathbb{P}_{\mathbb{X}_n^{[0,T]}}$ for all $n \in \mathcal{N}$, $\mathbb{P}^*_{\mathbb{B}^{[0,T]*}} = \mathbb{P}_{\mathbb{B}^{[0,T]}}$, and $\mathbb{P}^*\left(\lim_{n\rightarrow \infty}\mathbb{X}_n^{[0,T]*} =\mathbb{B}^{[0,T]*} \right) = 1$. Thus, we have
\begin{align*}
	\mathbb{S}_{[0,T]}^{(n)} &= \left\{S_t^{(n)} = S^{(n)}_{t_{n,k}},\;t\in[t_{n,k},t_{n,k+1}),\;k=0,\ldots,k_n-1; \;
S_{t_n,0}^{(n)} = S_0,\;S_T^{(n)} = S_{t_{n,k_n}}^{(n)}\right\},
	\\
	\mathbb{S}_{[0,T]} &= \left\{S_t = S_0 \exp \left\{\left(\mu-\frac{\sigma^2}{2}\right)t+\sigma B_t\right\},\;t\in[0,T]\right\},
\end{align*}
and $\mathbb{S}_{[0,T]}^{(n)}$ converges weakly to $\mathbb{S}_{[0,T]}$ in $(\mathcal{D}[0,T],d^{(0)})$. We note that in using DPIP2 the discrete pricing tree (\ref{eq5_b_tree}) depends on $\mu$ and $p_{n,k}$ while the continuous option price $\mathbb{S}_{[0,T]}$ only depends on $\mu$.

\subsection{Estimation of $\mathbf{p_{n,\Delta t}}$}
\label{sec23}

\begin{figure}
    \begin{center} 
    	\centering\includegraphics[width=0.5\textwidth]{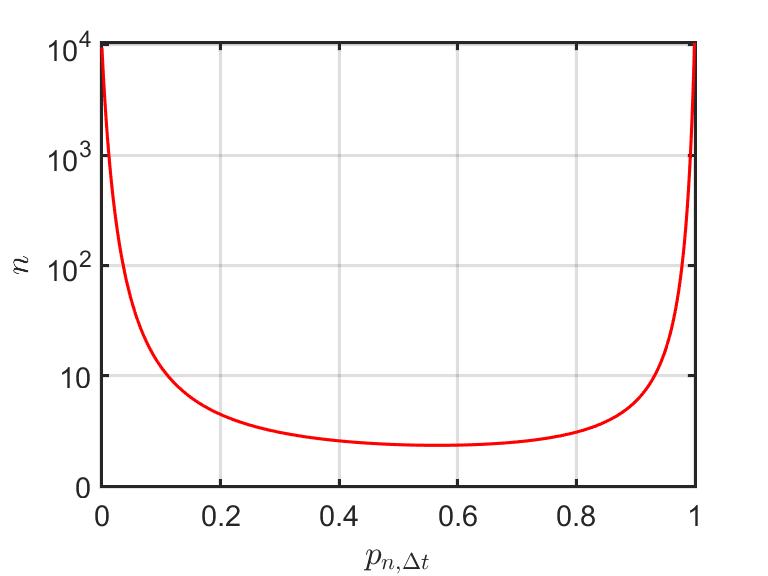}
	\caption{Convergence rate, measured as $n = T / \Delta t$, of  $S_T^{(n)}$ to $S_T^{(q)}$ as a function of $p_{n,\Delta t}$.
		$S_T^{(n)}=S_{t_{n,k_{n}}}^{(n)}$ is defined in (\ref{eq5_b_tree}) and $S_T$ in (\ref{eq1_gbm}).
		Here $T = 1$ and convergence was considered achieved when $|(S_T^{(n)}-S_T)/S_0 | \leq 10^{-6}$.}
    	\label{fig1_pvsn}
    \end{center}
\end{figure}

\noindent
Let $\Delta t = \Delta t_{n,k} = T/n$  and  $p_{n,k}=p_{n,\Delta t},\;k=1,\ldots,n$.
As shown in Section \ref{sec22}, $S_{[0,T]}^{(n)}$ converges weakly to $S_{[0,T]}$.\footnote{
For recent results on the rate of convergence in the central limit theorem see the survey in \cite{Senatov2017}.
Their results indicate the reason behind the rapid increase on $n$ when $p_{n,\Delta t}$ approaches 1 or 0.}
\cite{Kim2016} studied that the rate of convergence of $S_{[0,T]}^{(n)}$ to $S_{[0,T]}$.\footnote{
The theoretical bounds for the rate of convergence in DPIP have been the subject of numerous papers, see for example, \cite{Borisov1984}.
In our univariate $(t=T)$ case, the theoretical bounds of the rate of convergence are known as Berry-Esseen bounds;
see for example \cite{Chen2011}, Chapter 3.}
The rate of convergence deteriorates as $p_{n,\Delta t}$ approaches 1 or 0.
We illustrate the convergence issue with a numerical computation, the results of which are shown in Figure \ref{fig1_pvsn}.
Given $p_{n,\Delta t} \in (0,1)$, we compute the value of $n$ needed such that $\left | S_T^{(n)} - S_T \right | \le 10^{-4}$.
In the limit $(n\uparrow \infty,\Delta t \downarrow 0)$, the importance of $p_{n,\Delta t}$ vanishes since the trader taking a short position in
$\mathcal{C}$ is allowed to trade continuously in time (which, indeed, is a fiction in real trading).

\begin{figure}
 	\begin{center} 
    	\centering\includegraphics[width=400pt]{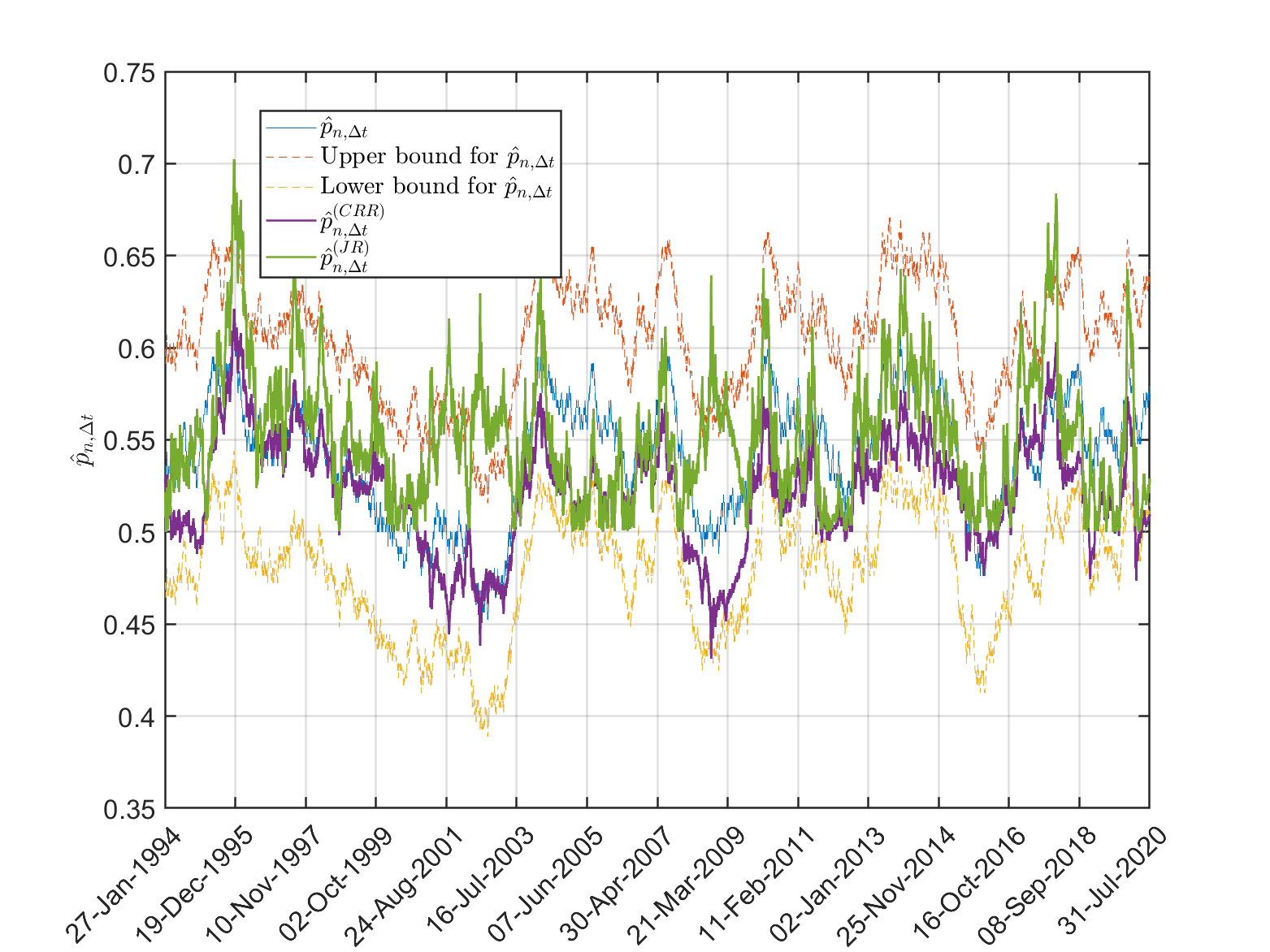}
    	\caption{Values of $\hat{p}_{n\Delta t}$ of the SPY daily log-returns computed from
		a one-year moving window over the time period $01/27/1994$ to $07/31/2020$.
		Also show are values for $\hat{p}_{n\Delta t}^{(\text{CRR})}$ and $\hat{p}_{n\Delta t}^{(\text{JR})}$
		computed as described in Section \ref{subsec_31}. }
    	\label{fig3_empiricalppp}
    \end{center}
\end{figure}

\begin{figure}
 	\begin{center} 
    	\centering\includegraphics[width=0.4\textwidth]{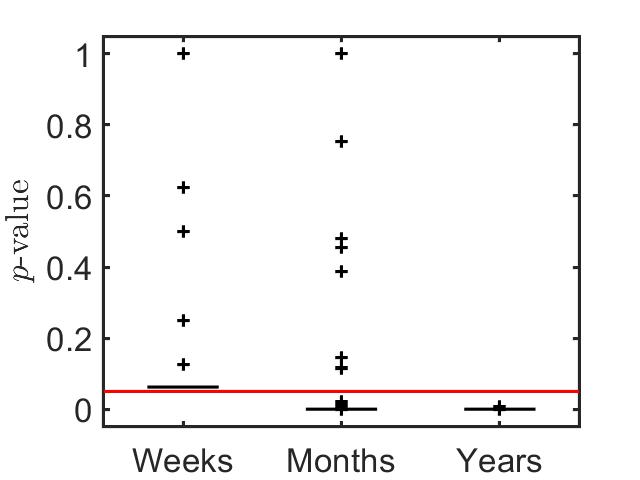}
    	\caption{Box-whisker plots of the $p$-values for two-sided sign tests on the values of $\hat{p}_{n\Delta t}$ of the SPY daily log-returns shown in Figure \ref{fig3_empiricalppp}.
		     The solid red line indicates the 0.05 significance level.}
    	\label{fig_pvalue_emp_vs_0.5}
    \end{center}
\end{figure}

We argue that an option trader should use the information incorporated in the estimates of the  probability $p_{n,\Delta t}$.
To emphasize this point, we provide estimates for $p_{n,\Delta t}$ based on 25.5 years of the SPDR S\&P 500 ETF (SPY)\footnote{
	See {\it https://finance.yahoo.com/quote/SPY?p=SPY\&.tsrc=fin-srch}.}
historical price data.
We use a window of one year to construct a moving window strategy to estimate 
\begin{equation*}
\hat{p}_{n,\Delta t} = \frac{\textup{number of days with non-negative log-return in the current window}}{\textup{total number of days in the current window}}.
\end{equation*}
The estimates for ${\hat{p}}_{n,\Delta t}$ are shown in Figure \ref{fig3_empiricalppp}.
We split the time series $\hat{p}_{n,\Delta t}$ into non-overlapping intervals (weeks, months, and years) and apply the two-sided sign test on each interval with
$H_0:\;\hat{p}_{n,\Delta t} = 1/2$.
Figure \ref{fig_pvalue_emp_vs_0.5} shows the results for the hypothesis test in terms of box-whisker plots of the $p$-values.
The plots: indicate there is insufficient evidence to reject $H_0$ at the $0.05$ significance level for weekly intervals;
imply rejection of $H_0$ in most monthly intervals; and imply rejection of $H_0$ in all yearly intervals.

\section{General Binomial Option Pricing}
\label{sec3}
\noindent
Following the framework of the CRR-binomial pricing model, our next goal is to use the general binomial tree (\ref{eq5_b_tree})
to derive the discrete price dynamics $f_{t_{n,k}}^{(n)},\;k=0,\ldots,k_{n-1}$ of an ECC having terminal payoff
$f_T^{(n)}=g(S_{t_{n,k_n}})$. 
For $k=0,1,\ldots,k_n-1$ consider the replicating risk-neutral portfolio
$P_{t_{n,k}}^{(n)} = D_{t_{n,k}}^{(n)}S_{t_{n,k}}^{(n)}-f_{t_{n,k}}^{(n)}$, $P_{t_{n,k+1}}^{(n)}
= D_{t_{n,k}}^{(n)}S_{t_{n,k+1}}^{(n)}-f_{t_{n,k+1}}^{(n)}$, with $D_{t_n,k}^{(n)}$ being the ``delta" position,
and $S_{t_{n,k+1}}^{(n)}$ determined by the pricing tree (\ref{eq5_b_tree}).
From the hedge position,
\begin{equation*}
D_{t_{n,k}}^{(n)}S_{t_{n,k+1}}^{(n,u)}-f_{t_{n,k+1}}^{(n,u)} = D_{t_{n,k}}^{(n)}S_{t_{n,k+1}}^{(n,d)}-f_{t_{n,k+1}}^{(n,d)},
\end{equation*} 
it follows
\begin{align}
	D^{(n)}_{t_{n,k}}  &= \frac{f^{(n,u)}_{t_{n,k+1}}-f^{(n,d)}_{t_{n,k+1}}}{S_{t_{n,k}}^{(n)}e^{\mu \Delta t_{n,k+1}}\left\{e^{M_1}-e^{M_2}\right\}},
	\label{eq6_delta}
	\end{align}
	where
	\begin{align}
	M_1 & = -\frac{1-p_{n,k+1}}{p_{n,k+1}}\frac{\sigma^2}{2}\Delta t_{n,k+1}+\sigma\sqrt{\frac{1-p_{n,k+1}}{p_{n,k+1}}}\sqrt{\Delta t_{n,k+1}},
	\nonumber
	\\
	M_2 &= -\frac{p_{n,k+1}}{1-p_{n,k+1}}\frac{\sigma^2}{2}\Delta t_{n,k+1}-\sigma\sqrt{\frac{p_{n,k+1}}{1-p_{n,k+1}}}\sqrt{\Delta t_{n,k+1}}.
	\nonumber
\end{align}
As $f_{t_{n,k}}^{(n)} = D_{t_{n,k}}^{(n)}S_{t_{n,k}}^{(n)}-e^{-r\Delta t_{n,k+1}}P_{t_{n,k+1}}^{(n,u)}$, we have
\begin{equation*}
	f_{t_{n,k}}^{(n)} = e^{-r\Delta t_{n,k+1}}\left\{q_{n,k+1}f_{t_{n,k+1}}^{(n,u)}+(1-q_{n,k+1})f_{t_{n,k+1}}^{(n,d)}\right\},
\end{equation*}
where the risk-neutral upturn probability $q_{n,k+1}$ for the time period $[t_{n,k},t_{n,k+1})$ is given by
\begin{equation}
	q_{n,k+1} = \frac{e^{(r-\mu)\Delta t_{n,k+1}}-e^{M_2}}{e^{M_1}-e^{M_2}},
	\label{eq7_q}
\end{equation}
with $M_1$ and $M_2$ defined in (\ref{eq6_delta}).
Assuming all terms of order $o(\Delta t_{n,k})$ are negligible in (\ref{eq7_q}), we have
\begin{equation}
	q_{n,k} = p_{n,k} - \theta\sqrt{p_{n,k}(1-p_{n,k})\Delta t_{n,k}},\; k=1,\ldots,k_n,\;n\in\mathcal{N},
	\label{eq8_qp}
\end{equation}
where $\theta = (\mu-r)/\sigma$ is the market price of risk.

\noindent
Furthermore, the delta-position $D^{(n)}_{t_{n,k}}$ in (\ref{eq6_delta}) becomes
\begin{equation*}
	D^{(n)}_{t_{n,k}} = \frac{\left(f^{(n,u)}_{t_{n,k+1}}-f^{(n,d)}_{t_{n,k+1}}\right)\sqrt{p_{n,k+1}(1-p_{n,k+1})}}{S_{t_{n,k}}^{(n)}e^{\mu \Delta t_{n,k+1}}\sigma \sqrt{\Delta t_{n,k+1}}}.
\end{equation*}
Thus, as $p_{n,k+1} \uparrow 1$ or $p_{n,k+1} \downarrow 0$, the delta-position $D^{(n)}_{t_{n,k}}\rightarrow 0$, since if $p_{n,k+1} = 0$ or $p_{n,k+1} = 1,\;k=0,1,\ldots,k_n$, the ECC becomes a fixed-income security and no hedging is needed.

We now consider the risk-neutral dynamics,
\begin{equation}
	S^{(n,q)}_{t_{n,k}} =
	\begin{cases}
	S^{(n,q,u)}_{t_{n,k}}&=S^{(n,q)}_{t_{n,k-1}}\exp\left\{\left(\mu- \frac{1-p_{n,k}}{p_{n,k}}\frac{\sigma^2}{2}\right)\Delta t_{n,k}+\sigma\sqrt{\frac{1-p_{n,k}}{p_{n,k}}}\sqrt{\Delta t_{n,k}} \right\},\;\textup{if}\;\varsigma_{n,k}^{(q)} = 1,
	\\
	S^{(n,q,d)}_{t_{n,k}}&=S^{(n,q)}_{t_{n,k-1}}\exp\left\{\left(\mu- \frac{p_{n,k}}{1-p_{n,k}}\frac{\sigma^2}{2}\right)\Delta t_{n,k}-\sigma\sqrt{\frac{p_{n,k}}{1-p_{n,k}}}\sqrt{\Delta t_{n,k}} \right\},\;\textup{if}\;\varsigma_{n,k}^{(q)} = -1,
	\end{cases}
	\label{eq9_b_tree_q}
\end{equation}
where $\varsigma_{n,k}^{(q)},\;k=1,\ldots,k_n$ are independent binary random variables with $\mathbb{P}(\varsigma^{(q)}_{n,k} =1) =1-\mathbb{P}(\varsigma^{(q)}_{n,k} =-1) = q_{n,k}\in (0,1),\; k =1,\ldots,k_n,\; n\in \mathcal{N}$. The discrete risk-neutral log-return process $R_{t_{n,k}}^{(n,q)} = \ln \left(S^{(n,q)}_{t_{n,k}}/S^{(n,q)}_{t_{n,k-1}}\right)$, has mean $\mathbb{E}\left(R_{t_{n,k}}^{(n,q)}\right) = \left(r-\frac{\sigma^2}{2}\right)\Delta t_{n,k}$ and variance $\textup{Var}\left(R_{t_{n,k}}^{(n,q)}\right) = \sigma^2 \Delta t_{n,k}$.  Set
\begin{align}
	\mathbb{S}_{[0,T]}^{(n,q)} &= \left\{S_t^{(n,q)} = S^{(n,q)}_{t_{n,k}},\;t\in[t_{n,k},t_{n,k+1}),\;k=0,\ldots,k_{n-1},\;S_{t_n,0}^{(n,q)} = S_0,\;S_T^{(n,q)} = S_{t_{n,k_n}}^{(n,q)}\right\},
	\nonumber
	\\
	\mathbb{S}^{(q)}_{[0,T]} &= \left\{S^{(q)}_t = S_0 \exp \left\{\left(r-\frac{\sigma^2}{2}\right)t+\sigma B^{(q)}_t\right\},\;t\in[0,T]\right\},
	\label{eq10_S_q}
\end{align}
where $\mathbb{B}_{[0,T]}^{(q)} = \left\{B^{(q)}_t\right\}_{0\leq t\leq T}$ is a standard BM.
Then, using the same arguments as in Section \ref{sec22} we have that $\mathbb{S}_{[0,T]}^{(n,q)}$ converges weakly to
$\mathbb{S}^{(q)}_{[0,T]}$ in $(\mathcal{D}[0,T],d^{(0)})$.

\subsection{Comparison with CRR and JR models}
\label{subsec_31}

\begin{figure}
\begin{center}
    \begin{subfigure}[b]{0.4\textwidth} 
    	\includegraphics[width=\textwidth]{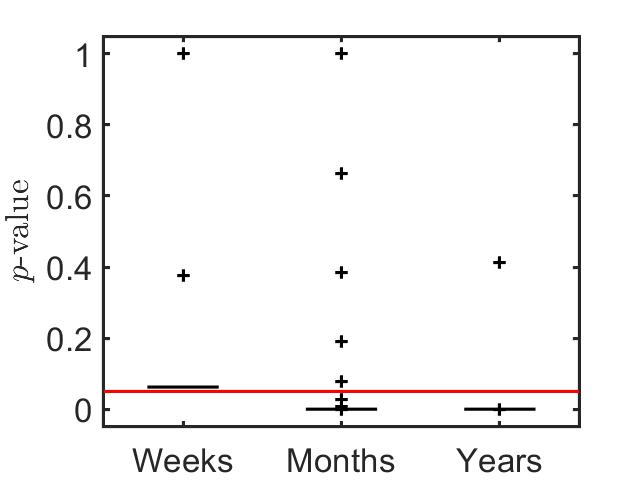}
    	\caption{CRR-model.}
    	\label{fig_pvalue_empCRR}
    \end{subfigure}
    \begin{subfigure}[b]{0.4\textwidth} 
    	\includegraphics[width=\textwidth]{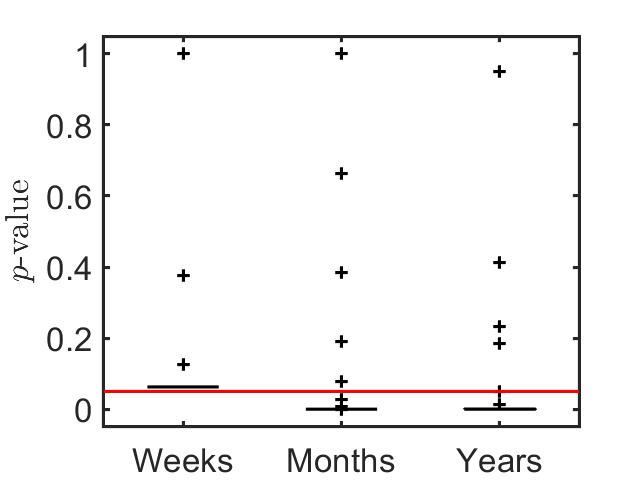}
    	\caption{JR-model.}
    	\label{fig_pvalue_empJR}
    \end{subfigure}
    \caption{Box-whisker plots of the $p$-values for two-sided sign tests on the values of $\hat{p}_{n\Delta t}^{(\text{CRR})}$ and $\hat{p}_{n\Delta t}^{(\text{JR})}$
		   shown in Figure \ref{fig3_empiricalppp}.
		   The solid red line indicates the 0.05 significance level.}
\end{center}
\end{figure}

\noindent
If $p_{n,k} = p_{n,\Delta t} = p^{(0)}+p^{(1)}\sqrt{\Delta t}+p^{(2)}\Delta t,\; k=1,\ldots,k_n,\; \Delta t = T/n$,
then from (\ref{eq8_qp})
\begin{equation*}
	q_{n,k} = q_{n,\Delta t} \vcentcolon
	= p^{(0)}
	+ \left(p^{(1)} - \theta\sqrt{p^{(0)}(1-p^{(0)})}\right)\sqrt{\Delta t}
	+ \left(p^{(2)}-\frac{\theta}{2} \frac{p^{(1)}(1-2p^{(0)})}{\sqrt{p^{(0)}(1-p^{(0)})}}\right)\Delta t.
\end{equation*}
In the CRR-model, the risk-neutral probability for non-negative stock log-returns in period
$k\Delta t,\;\Delta t= \frac{T}{n},\;n\uparrow \infty$ is given by
$q_{n,k}^{(\text{CRR})} = q_{n,\Delta t}^{(\text{CRR})}
= \frac{\exp(r\Delta t) - \exp(-\sigma\sqrt{\Delta t})}{\exp(\sigma\sqrt{\Delta t}) - \exp(-\sigma\sqrt{\Delta t})}
= 1/2+\frac{r - \sigma^2/2}{2\sigma} \sqrt{\Delta t}$.
According to (\ref{eq8_qp}), the corresponding natural probability $p_{n,k}^{(\text{CRR})} = p_{n,\Delta t}^{(\text{CRR})}$ is given by
$p_{n,\Delta t}^{(\text{CRR})}=1/2+\frac{\mu-\sigma^2/2}{2\sigma}\sqrt{\Delta t}$.
In the JR-model, $q_{n,k}^{(\text{JR})} = q^{(\text{JR})}=1/2$.
If $\Delta t=\frac{T}{n}$ then as $n \uparrow \infty$,
the corresponding probability for stock-upturn in the JR-model is given by
$p_{n,k}^{(\text{JR})} = p_{n,\Delta t}^{(\text{JR})}=1/2+|\theta|\sqrt{\Delta t}$.
Figure \ref{fig3_empiricalppp} also plots estimates of the values for  $\hat{p}_{\Delta t}^{(\text{CRR})}$and $\hat{p}_{\Delta t}^{(\text{JR})}$
using the SPY data of Section \ref{sec23}, comparing them with the estimated values for $\hat{p}_{\Delta t}$.
Again, we apply a two-sided sign test with the following hypotheses on non-overlapping intervals (weeks, months, and years):
\begin{equation*}
\begin{cases}
H_0^{(\text{CRR})}:\;\hat{p}_{\Delta t}=\hat{p}_{\Delta t}^{(\text{CRR})}, \\
H_a^{(\text{CRR})}:\;\hat{p}_{\Delta t}\neq\hat{p}_{\Delta t}^{(\text{CRR})}, \\
\end{cases} \qquad
\begin{cases}
H_0^{(\text{JR})}:\;\hat{p}_{\Delta t}=\hat{p}_{\Delta t}^{(\text{JR})}, \\
H_a^{(\text{JR})}:\;\hat{p}_{\Delta t}\neq\hat{p}_{\Delta t}^{(\text{JR})}. \\
\end{cases}
\end{equation*}
The results from the two-sided sign tests are shown as box-whisker plots in Figures \ref{fig_pvalue_empCRR} and \ref{fig_pvalue_empJR}.
They produce conclusions somewhat similar to those of Figure \ref{fig_pvalue_emp_vs_0.5}.
In the case of weekly intervals in both figures, there is no sufficient evidence to reject $H_0^{(\text{CRR})}$ ($H_0^{(\text{JR})}$) at the $0.05$ significance level,
whereas for monthly and year intervals, the results indicate rejection of $H_0^{(\text{CRR})}$ ($H_0^{(\text{JR})}$) in most intervals.

\subsection{Rate of loss of $\mu$ as hedging rate increases}
\label{subsec_32}

\begin{figure}
 	\begin{center} 
    	\centering\includegraphics[width=0.4\textwidth]{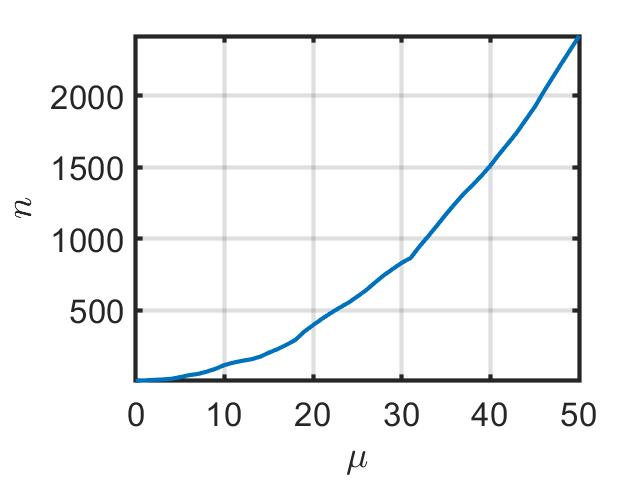}
    	\caption{Convergence rate, measured as $n = T / \Delta t$, of  $S_T^{(n,q)}$ to $S_T^{(q)}$ in (\ref{eq10_S_q}) as a function of $\mu$ over the range $\mu\in(0,50)$.
	 In this simulation, $r = 0$, $p_{n,\Delta t} = 0.5$, $\sigma = 1$, $T = 1$ and convergence was considered achieved when
	$|(S_T^{(n,q)}-S_T^{(q)})/S_0 | \leq 10^{-6}$.}
    	\label{fig4_mun}
    \end{center}
\end{figure}

In continuous-time option pricing, the mean log-return parameter $\mu$  and the information about the market direction embedded in $p_{n,k}$ are lost
due to the artificial assumption that hedging can be done continuously in time with no transaction costs.
We can estimate the rate of loss of $\mu$ from (10) by computing the dependence of the number of necessary hedging instances, $n = T / \Delta t$ on $\mu$ while requiring that $\| S_T^{(n,q)} - S_T^{(q)} \| \le 10^{-4}$.
The results over the range $\mu \in [0,50]$ are shown in Figure \ref{fig4_mun}.
From the figure we deduce that $n \uparrow \infty$ as $\mu \uparrow \infty$.
From these results we conclude that, in continuous option pricing, $\mu$ disappears as a model parameter due to the (practically inconceivable) use of continuous time hedging.

\subsection{Estimation of the implied risk-neutral probability $\mathbf{q_{n,\Delta t}}$}

\noindent
We estimate the risk-neutral probability $q_{n,\Delta t}$ in (\ref{eq8_qp}) based on XSP.\footnote{
	The CBOE Mini-SPX option contract, known by its symbol XSP, is an index option product designed to track
	the underlying S\&P 500 Index.
	See {\it http://www.cboe.com/products/stock-index-options-spx-rut-msci-ftse/s-p-500-index-options/}.}
Denote the call option prices by $C^{(\text{XSP})}_t(K,T-t)$, where $K$ is the strike price, and $T$ is the terminal time.
We assume that hedging occurs daily, that is, $\Delta t_{n,k} = k\Delta t,\;k = 1,...,n,\;n = T/\Delta t$.
Then, (\ref{eq8_qp}) becomes $q_{n,\Delta t} = p_{n,\Delta t} - \theta\sqrt{p_{n,\Delta t}(1-p_{n,\Delta t})\Delta t}$
and through terms of $O(\Delta t)$,
\begin{equation}
p_{n,\Delta t} = q_{n,\Delta t}+|\theta|\sqrt{q_{n,\Delta t}\left(1-q_{n,\Delta t}\right)\Delta t}+\left(\frac{1}{2}-q_{n,\Delta t}\right)\theta^2\Delta t.
\label{eq99_newpq}
\end{equation}
The XSP call options data were collected on $07/31/2020$ with initial capital $S_0^{(\text{SPY})} = 326.52$ and annual risk-free rate\footnote{
Here, use 10-year Treasury rate as risk-free rate,
see {\it https://www.treasury.gov/resource-center/data-chart-center/interest-rates/pages/textview.aspx?data=yield}.} $r = 0.55\%$.
We estimated $\mu$ and $\sigma$ using the mean and standard derivation of the log-return of SPY for the one-year window from $08/02/2019$ to $07/31/2020$.
This window produced the estimates $\hat{\mu} = 6.20\times 10^{-4}$ and $\hat{\sigma} = 0.02$.
Following the framework established in Section \ref{sec3} and (\ref{eq99_newpq}),
we can view the theoretical value of the call options
$C^{(\text{XSP;theoretical})}_t(S^{(\text{SPY})}_0,K,T-t,q_{n,\Delta t})$ as a function of $q_{n,\Delta t}$.
For different strike price $K$ and time $t$, we can estimate $q_{n,\Delta t}$ via
\begin{equation*}
	\hat{q}_{n,\Delta t} =
	 \textup{arg\;min} \left\{\left(\frac{C^{(\text{XSP;theoretical})}_t(S^{(\text{SPY})}_0,K,T-t,q_{n,\Delta t})-C^{(\text{XSP})}_t(K,T-t)}{C^{(\text{XSP})}_t(K,T-t)}\right)^2\right\}.
\end{equation*}

\begin{figure}[ht]
    \centering
    \subcaptionbox{$\hat{q}_{n,\Delta t}$}{\includegraphics[width=0.32\textwidth]{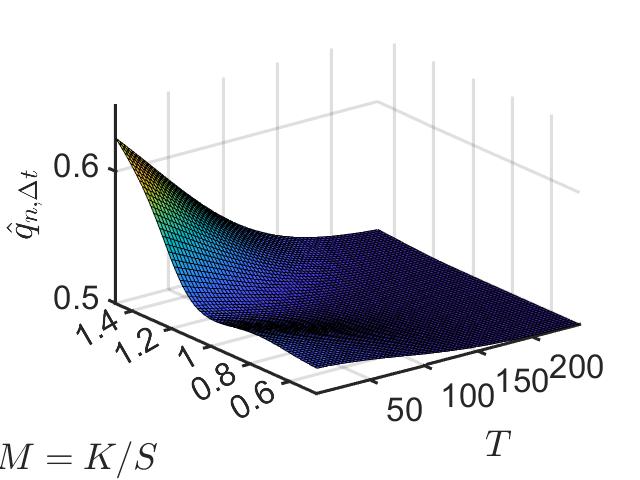}}\hspace{0em}%
    \subcaptionbox{$\hat{p}_{n,\Delta t}$}{\includegraphics[width=0.32\textwidth]{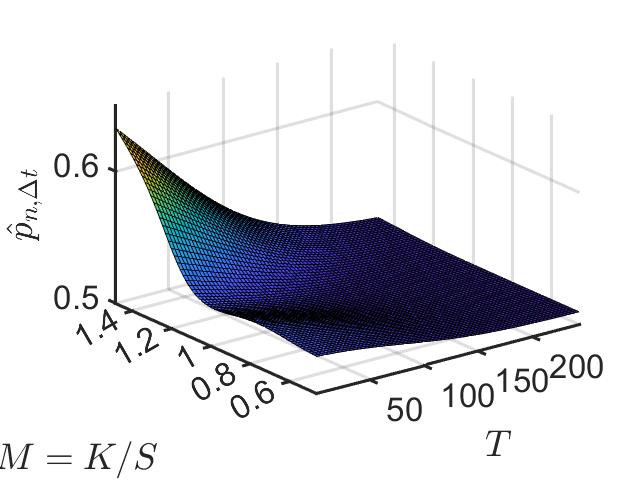}}\hspace{0em}%
    \subcaptionbox{$\hat{q}_{n,\Delta t} - \hat{p}_{n,\Delta t}$}{\includegraphics[width=0.32\textwidth]{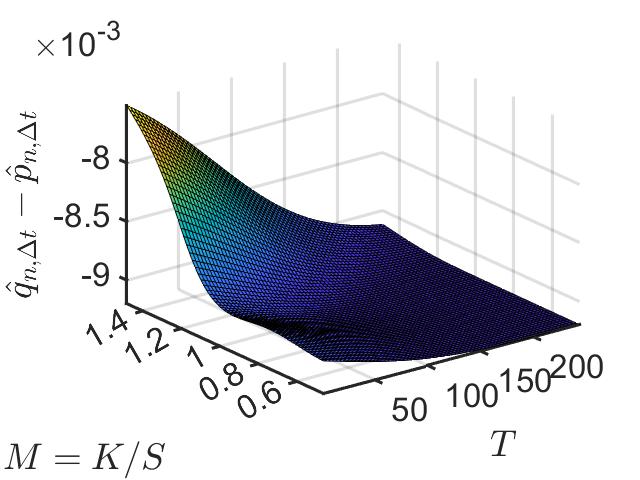}}\hspace{0em}%
    \caption{Implied surfaces for $\hat{q}_{n,\Delta t}$, $\hat{p}_{n,\Delta t}$, and $\hat{q}_{n,\Delta t} - \hat{p}_{n,\Delta t}$ plotted as functions of time to
	maturity $T$ and moneyness $M=K/S$.}
    \label{fig6_pq_minus}
\end{figure}

Figures \ref{fig6_pq_minus}(a-c) present the implied $\hat{q}_{n,\Delta t}$, the corresponding implied $\hat{p}_{n,\Delta t}$, and $\hat{q}_{n,\Delta t}-\hat{p}_{n,\Delta t}$ surfaces. All figures are graphed against the standard measures of moneyness and time to maturity (in days).
From Figure \ref{fig6_pq_minus}(a),  $\hat{q}_{n,\Delta t}$ ranges from 0.5 to 0.62.
Given these values of $\hat{q}_{n,\Delta t}$ and values of $r$, $\sigma$ and $\mu$, equation (\ref{eq99_newpq}) shows that
$\hat{p}_{n,\Delta t}$ = $\hat{q}_{n,\Delta t} + O(10^{-3})$.\footnote{
	This reflects the desire of a central bank not to disturb the tendencies of the real world by too large a factor.
	See e.g. \cite{Dhaene2015}, \cite{Esche2005} and \cite{Fujiwara2003} for related work on minimal entropy martingale
	measures applied to markets.}
	
As a result, Figure \ref{fig6_pq_minus}(b) shows that $\hat{p}_{n,\Delta t}$ varies from 0.51 to 0.63.
For a fixed $T$, $\hat{p}_{n,\Delta t}$ has a higher value as $M\in (1,1.5)$ compared to when $M\in (0.5,1)$.
And for any value of $M$, $\hat{p}_{n,\Delta t}$ decreases as time to maturity increases.
Recall that $p_{n,\Delta t}$ is the natural probability for an upward movement (or non-negative log-return) of the stock price over the time period
$\Delta t$, and $q_{n,\Delta t}$ is the corresponding risk-neutral probability for an upward movement of the stock price.
Thus the implied surface indicates that, on the trading day $07/31/2020$, the option traders of SPY were optimistic (bullish) for the coming six-month period.

\section{Cherny-Shiryaev-Yor Invariance Principle and Path-dependent Stock Log-return Dynamics}
\label{sec4}

\noindent 
In this section we formulate a new binomial path-dependent pricing model where the stock price dynamics depends on the log-return
dynamics of a market index.
Our approach is based on an extension of the Donsker-Prokhorov invariance principle due to \cite{Cherny2003}. 
We start with the formulation of the Cherny-Shiryaev-Yor invariance principle (CSYIP).
Let $\xi_k,\;k\in \mathcal{N}$ be a sequence of independent and identically distributed ({\it i.i.d.}) random variables with mean 0 and variance 1.
Set $\xi_0^{(n)} = 0,\;\xi_{k}^{(n)} = n^{-\frac{1}{2}}\xi_k$,
and $X_{k/n}^{(n)} = \sum_{i=1}^n\xi_{i}^{(n)},\; k\in \mathcal{N},\; n\in \mathcal{N}$.
Let $\mathbb{B}^{(n)}_t,\; t \geq 0$ be the random process with piecewise linear trajectories having vertices $(k/n,\mathbb{B}^{(n)}_{k/n})$,
where $\mathbb{B}^{(n)}_{k/n} = X^{(n)}_{k/n}, \; k \in \mathcal{N},\; n \in \mathcal{N}$.
Following \cite{Cherny2003}, call a function $h:R \rightarrow R$ a {\it CSY piecewise continuous function} if there exists
a collection of disjoint intervals $J_n,\;n\in \mathcal{N}$\footnote{
Each $J_n$ can be a closed, open, semi-open interval or a point.} such that:
\begin{itemize}
	\item[(i)]$\cup^{\infty}_{n=1}J_n = R$;
	\item[(ii)]for every compact interval $J$, there exists $n \in \mathcal{N}$ such that $\cup^{n}_{k=1}J_k \supseteq J$;
	\item[(iii)]on each $J_n,\; n\in \mathcal{N}$, the function $h: J_n \rightarrow R$ is continuous, and has finite limits at those endpoints of $J_n$ which do not belong to $J_n$. 
\end{itemize}
Let $Y_{k/n}^{(n)} = \sum_{i=1}^k h\left(X^{(n)}_{(i-1)/n}\right)\xi_{i}^{(n)},\; k\in \mathcal{N},\; n\in \mathcal{N}$.
For every fixed $n\in\mathcal{N}$, define $\mathbb{C}^{(n)}_t,\; t \geq 0$ to be the random process with CSY piecewise linear trajectories having
vertexes $(k/n,\mathbb{C}^{(n)}_{k/n})$, where $\mathbb{C}^{(n)}_{k/n} = Y^{(n)}_{k/n}, \; k \in \mathcal{N},\; n \in \mathcal{N}$.

\noindent({\bf CSYIP}) {\it If $h:R\rightarrow R$ is a CSY piecewise continuous function, then, as $n\uparrow \infty$,
the bivariate process $(\mathbb{B}_t^{(n)},\mathbb{C}_t^{(n)}),\; t\geq 0$ converges in law\footnote{
For the definition of convergence in law, see Section 2.2, page 45 of \cite{Jarod2012}.}
to $(B_t,C_t),\; t\geq 0$, where $B_t$ is a standard BM and $C_t = \int_0^t h(B_s)dB_s$.} 

In the CRR- and JR-binomial models, as well as in the general binomial model (\ref{eq5_b_tree}), the stock log-returns are assumed independent.
Based on CSYIP, we introduce a binomial tree model in which the log-returns are dependent on a sequence of random signs representing the past
history of the movement of a market index influencing the stock's dynamics.
Let $S_{t_{n,k}}^{(n,M)},\;t_{n,k}=k\Delta t,\;k = 1,\ldots,n,\;n\Delta t= T,\;n\in N$, with $S_{t_{n,0}}^{(n,M)} = S_0^{(n,M)} = S_0^{(M)} > 0$,
be a market index value for the period $[t_{n,k},t_{n,k+1})$.
Let $R_{t_{n,k}}^{(n,M,hist)}=\ln \left(S_{t_{n,k}}^{(n,M,hist)}/S_{t_{n,k-1}}^{(n,M,hist)}\right)$ be the historical market index log-return (index-return for brevity) in the
$k^{th}$ period, and $\varsigma_{n,k}^{(M)},\;k=1,\ldots,n$, be the binary sequence of the index’s value directions:
$\varsigma_{n,k}^{(M)}=1$, if $R_{t_{n,k}}^{(n,M,hist)} \geq 0$, and $\varsigma_{n,k}^{(M)}= -1$, if $R_{t_{n,k}}^{(n,M,hist)}<0$.
We assume that $\varsigma_{n,k}^{(M)},\;k=1,\ldots,n$, are independent random signs,
with $\mathbb{P}(\varsigma^{(M)}_{n,k} =1) =1-\mathbb{P}(\varsigma^{(M)}_{n,k} =-1) = p^{(M)}_{n,\Delta t}
= p^{(M)}_{0}+p^{(M)}_{1}\sqrt{\Delta t}+p^{(M)}_{2}\Delta t$.
We further assume $p^{(M)}_{0}\in (0,1),\;p^{(M)}_{1}\in R,\; p^{(M)}_{2} \in R,\; \Delta t = \frac{T}{n}\downarrow 0$ so that $p_{n,\Delta t}^{(M)} \in (0,1)$.
We model the market index price dynamics (index-dynamics for brevity) as
\begin{equation}
	S_{t_{n,k}}^{(n,M)} = S_{t_{n,k-1}}^{(n,M)}\exp\left\{R_{t_{n,k}}^{(n,M)}\right\},
	\label{eq11_S}
\end{equation}
where
\begin{equation}
	R_{t_{n,k}}^{(n,M)} = 
	\begin{cases}
	\left(\mu^{(M)}- \frac{1-p^{(M)}_{n,\Delta t}}{p^{(M)}_{n,\Delta t}}\frac{{\sigma^{(M)}}^2}{2}\right)\Delta t+\sigma^{(M)}\sqrt{\frac{1-p^{(M)}_{n,\Delta t}}{p^{(M)}_{n,\Delta t}}}\sqrt{\Delta t},\;\textup{if}\;\varsigma_{n,k}^{(M)} = 1,
	\\
	\left(\mu^{(M)}- \frac{p^{(M)}_{n,\Delta t}}{1-p^{(M)}_{n,\Delta t}}\frac{{\sigma^{(M)}}^2}{2}\right)\Delta t-\sigma^{(M)}\sqrt{\frac{p^{(M)}_{n,\Delta t}}{1-p^{(M)}_{n,\Delta t}}}\sqrt{\Delta t} ,\;\textup{if}\;\varsigma_{n,k}^{(M)} = -1,
	\end{cases}
	\label{eq12_R}
\end{equation}
with $\mu^{(M)}>r>0$, and $\sigma^{(M)}>0$ determined from the historical log-return data.
As shown in Section \ref{sec21}, if 
\begin{align*}
	\mathbb{S}_{[0,T]}^{(n,M)} &= \left\{S_t^{(n,M)} = S^{(n,M)}_{t_{n,k}},\;t\in[t_{n,k},t_{n,k+1}),\;k=0,\dots, n-1;
\quad S_{t_n,0}^{(n,M)} = S_0,\;S_T^{(n,M)} = S_{t_{n,n}}^{(n,M)}\right\},
	\\
	\mathbb{S}^{(M)}_{[0,T]} &= \left\{S^{(M)}_t = S_0 \exp \left\{\left(\mu^{(M)}-\frac{{\sigma^{(M)}}^2}{2}\right)t+\sigma^{(M)} B_t\right\},\;t\in[0,T]\right\},
	\nonumber
\end{align*}
then the discrete market index price dynamics $\mathbb{S}_{[0,T]}^{(n,M)}$ converge weakly to $\mathbb{S}^{(M)}_{[0,T]}$ in $(\mathcal{D}[0,T],d^{(0)})$.

We apply CSYIP by setting
\begin{align}
&(a)\;Z^{(M)}_{n,k} = \frac{1}{\sigma^{(M)}\sqrt{\Delta t}}\left(R^{(n,M)}_{t_{n,k}}-\left(\mu^{(M)} - \frac{{\sigma^{(M)}}^2}{2}\right)\Delta t\right),
\nonumber
\\
&(b)\;\xi^{(M)}_{n,k} =
 \sqrt{\frac{1-\mathbbm{p}^{(M)}_{n,\Delta t}}{\mathbbm{p}^{(M)}_{n,\Delta t}}}I_{\left\{Z^{(M)}_{n,k} \geq 0 \right\}}
-
 \sqrt{\frac{\mathbbm{p}^{(M)}_{n,\Delta t}}{1-\mathbbm{p}^{(M)}_{n,\Delta t}}}I_{\left\{Z^{(M)}_{n,k} < 0 \right\}},
\label{eq13_setting_CSYip}
\\
&(c)\;X_{k/n}^{(n)} = \sum_{i=1}^k\sqrt{\Delta t}\xi_{n,i}^{(M)},
\nonumber
\\
&(d)\;Y_{k/n}^{(n,h)} = \sum_{i=1}^k\sqrt{\Delta t}\xi_{n,i}^{(M)}h\left(X_{(i-1)/n}^{(n)}\right),
\nonumber
\end{align}
where $\mathbbm{p}^{(M)}_{n,\Delta t} = \mathbb{P}\left(Z^{(M)}_{n,k} \geq 0\right) \in(0,1)$
is  the probability for an upturn in the index's centralized log-return $Z_{n,k}^{(M)}$.
The random signs $\xi_{n,k}^{(M)},\;k=1,\ldots,n$, are independent with
$\mathbb{E}\left(\xi_{n,k}^{(M)}\right)=0$ and $\textup{Var}\left(\xi_{n,k}^{(M)}\right) = 1$.
Consider the following processes in the Skorokhod space $\mathcal{D}[0,T]$, 
\begin{align}
	\mathbb{B}^{(n)}_{[0,T]} &= \left\{ B^{(n)}_t = X_{k/n}^{(n)},\; t\in [t_{n,k},t_{n,k+1}),\;k = 0, \ldots, n-1; \;
B_0^{(n)} = X_0^{(n)} = 0, \;B^{(n)}_T = X^{(n)}_1\right\},
	\label{eq14_BC}
	\\
	\mathbb{C}^{(n,h)}_{[0,T]} &= \left\{ C^{(n,h)}_t = Y_{k/n}^{(n,h)},\; t\in [t_{n,k},t_{n,k+1}),k = 0, \ldots, n-1; \;
C_0^{(n,h)} = Y_0^{(n,h)} = 0, \;C^{(n,h)}_T = Y^{(n,h)}_1\right\}.
	\nonumber
\end{align}
According to CSYIP, as $n\uparrow \infty$, the bivariate process $\left(\mathbb{B}_{[0,T]}^{(n)},\mathbb{C}_{[0,T]}^{(n,h)}\right)$
converges weakly in $\mathcal{D}[0,T]\times \mathcal{D}[0,T]$ to $\left(\mathbb{B}_{[0,T]},\mathbb{C}_{[0,T]}^{(h)}\right)$,
where $\mathbb{B}_{[0,T]}=\left\{B_t\right\}_{0\leq t\leq T}$ is a BM on $[0,T]$,
and $\mathbb{C}_{[0,T]}^{(h)}=\left\{C_t^{(h)}=\int_0^t h\left(B_s\right)d B_s\right\}_{0\leq t\leq T}$.
 
Next, we define the stock price discrete dynamics as a functional of $\mathbb{B}_{[0,T]}^{(n)}$ and $\mathbb{C}_{[0,T]}^{(n,h)}$.
Let
\begin{equation}
    \begin{aligned}
	\mathbb{S}_{[0,T]}^{(n,h,S)} ={} & \left\{ S_t^{(n,h,S)} = S_0^{(S)} \exp \left[ \nu t_{n,k}+\sigma B_t^{(n)}+ \gamma C^{(n,h)}_t \right], t \in [t_{n,k},t_{n,k+1}),\; k=0,\ldots,n-1, \right. \\
		& \left.\   S_T^{(n,h,S)} = S_0^{(S)}\exp \left[ \nu T+\sigma B_T^{(n)}+ \gamma C^{(n,h)}_T \right] \right\},
		\label{eq15_Snhs}
    \end{aligned}
\end{equation}
where $\nu>r>0,\;\sigma>0,\gamma\geq 0$ are parameters determining the dynamics of the stock price as a function of the index dynamics.
Then, as $n\uparrow \infty$, $\mathbb{S}_{[0,T]}^{(n,h,S)}$ converges weakly to
$\mathbb{S}_{[0,T]}^{(h,S)} = \left\{S_0^{(S)} \exp \left\{\nu t+\sigma B_t+ \gamma C^{(h)}_t\right\},\; t\in[0,T]\right\}$ in $\mathcal{D}[0,T]$.
The stock discrete log-return dynamics is given by
\begin{equation}
	R_{t_{n,k}}^{(n,h,S)} = \ln \left(\frac{S_{t_{n,k}}^{(n,h,S)}}{S_{t_{n,k-1}}^{(n,h,S)}}\right) = \nu \Delta t+\sigma \sqrt{\Delta t}\xi^{(M)}_{n,k}+ \gamma \sqrt{\Delta t}\xi^{(M)}_{n,k} h\left(\sum_{i=1}^{k-1}\sqrt{\Delta t}\xi^{(M)}_{n,i}\right).
	\label{eq16_Rnhs}
\end{equation}
Thus, when $\gamma \neq 0$, the stock log-return $R_{t_{n,k}}^{(n,h,S)}$ depends on  the entire path $\left\{ \xi_{n,i}^{(M)}, \; i=1,\ldots,k\right\}$ of market return intensity.
As a result, the $R_{t_{n,k}}^{(n,h,S)},\;k=1,\ldots,n$, are dependent log-returns when $\gamma \neq 0$.

\subsection{ An extension of the Cherny-Shiryaev-Yor invariance principle}
\label{sec43}

\noindent
In this section, we extend CSYIP to obtain a more flexible model for the stock price.
We extend (\ref{eq13_setting_CSYip}a-d) by adding the term,
\begin{align*}
(e)\;V_{k/n}^{(n,g)} = \sum_{i=1}^k\sqrt{\Delta t}\xi_{n,i}^{(M)}g\left(\sum_{j=1}^{i-1}\left(\sum_{l=1}^{j-1}\sqrt{\Delta t}\xi_{n,l}^{(M)}\right)\Delta t\right),
\end{align*}
where $g:R \rightarrow R$ is a CSY piecewise continuous function.
Together with the processes $\mathbb{B}_{[0,T]}^{(n)}$ and $\mathbb{C}_{[0,T]}^{(n,h)}$ in (\ref{eq14_BC}), we consider the additional $\mathcal{D}[0,T]$-process,
\begin{equation*}
		\mathbb{G}^{(n,g)}_{[0,T]} = \left\{ G^{(n,h)}_t = V_{k/n}^{(n,g)},\; t\in [t_{n,k},t_{n,k+1}),\;k = 0, \ldots, n-1,\;G_0^{(n,h)} = V_0^{(n,g)} = 0,\;G^{(n,g)}_T = V^{(n,g)}_1\right\},
\end{equation*}
and define the stock price discrete dynamics as a functional of $\mathbb{B}_{[0,T]}^{(n)}$, $\mathbb{C}_{[0,T]}^{(n,h)}$, and $\mathbb{G}^{(n,g)}_{[0,T]}$ by
\begin{equation}
    \begin{aligned}
	\mathbb{S}_{[0,T]}^{(n,h,g,S)} ={} & \left\{ S_t^{(n,h,g,S)} = S_0^{(S)} \exp \left[ \nu t_{n,k}+\sigma B_t^{(n)}+ \gamma C^{(n,h)}_t+\delta G_t^{(n,g)} \right],\; t\in[t_{n,k},t_{n,k+1}),\; k=0,\ldots,n-1, \right. \\
		& \left. \ S_T^{(n,h,g,S)} = S_0^{(S)} \exp \left[ \nu T+\sigma B_T^{(n)}+ \gamma C^{(n,h)}_T+\delta G_T^{(n,g)} \right] \right \},
	\label{eq17_Snhgs}
    \end{aligned}
\end{equation}
where $\nu>r>0,\;\sigma>0,\gamma> 0,\; \delta>0$.
As for the CSYIP discrete process, $\mathbb{S}_{[0,T]}^{(n,h,g,S)}$ converges weakly to 
\begin{equation}
	\mathbb{S}_{[0,T]}^{(h,g,S)} = \left\{ S_t^{(h,g,S)} = S_0^{(S)} \exp \left[ \nu t+\sigma B_t+ \gamma C^{(h)}_t+\delta G_t^{(g)} \right],\; t\in[0,T], \; S_0^{(S)}>0, \right\} 
	\label{eq18_Sconti}
\end{equation}
in $\mathcal{D}[0,T]$ as $n\uparrow \infty$, where: $B_t$ is a standard BM; $C_{t}^{(h)} = \int_0^t h(B_{\nu})dB_{\nu}$;  and $G_{t}^{(g)} = \int_0^t g(\int_0^{\nu} B_{u}du)dB_{\nu}$.
The stock discrete log-return dynamics is given by
\begin{align}
	&R_{t_{n,k}}^{(n,h,g,S)} = \ln \left(\frac{S_{t_{n,k}}^{(n,h,g,S)}}{S_{t_{n,k-1}}^{(n,h,g,S)}}\right) 
	\nonumber
	\\
	&= \nu \Delta t
	 + \sigma \sqrt{\Delta t}\xi^{(M)}_{n,k}
	 + \gamma \sqrt{\Delta t}\xi^{(M)}_{n,k} h\left(\sum_{i=1}^{k-1}\sqrt{\Delta t}\xi^{(M)}_{n,i}\right)
	 + \delta \sqrt{\Delta t}\xi_{n,k}^{(M)}g\left(\sum_{j=1}^{k-1}\left(\sum_{l=1}^{j-1}\sqrt{\Delta t}\xi_{n,l}^{(M)}\right)\Delta t\right).
	\label{eq19_Rnhgs}
\end{align}

The new term captures additional long-range dependence in $R_{t_{n,k}}^{(n,h,g,S)}$.
While the argument of the $h()$ function on the right-hand side of (\ref{eq19_Rnhgs}) weights past terms of $\xi^{(M)}_{n,i}$ equally in the summation,
the double summation in the argument of the $g()$ function has the effect of linearly weighting the $\xi^{(M)}_{n,i}$, with the weight decreasing toward
present time.\footnote{
The summations can be changed to have the weights linearly increasing toward present time without affecting any of the convergence properties of the model.}
Of course the (potentially non-linear) CSY piecewise continuous function $g()$ plays a role in mutating the long-range dependence introduced by its argument.
The Appendix to this paper provides a deeper numerical investigation of the behavior of the terms in (\ref{eq19_Rnhgs}) using MSFT for the stock and S\&P500
for the market index.
In particular it is shown that the choice of Gaussian functional forms for $h()$ and $g()$ act as band-pass filters on the informational content of their arguments.
Using a series of non-overlapping one-year time periods, the predictive ability of the model (\ref{eq19_Rnhgs}) is examined.
Finally the behavior of the discrete form (\ref{eq19_Rnhgs}) is compared to its continuum analog (\ref{eq18_Sconti}).

\subsection{An example}
\label{sec44}

\noindent
As an example of stock return dynamics depending on market intensity path, we consider the daily price process $S_{k\Delta t}^{(n,\text{MSFT})}$ of MSFT as a function
of the trajectory of the S$\&$P index dynamics.
MSFT and S\&P 500 price data for the period $07/01/2019$ through $06/30/2020$ were obtained from Yahoo Finance.
From this period we estimated the mean $\mu^{(\text{S\&P})}$ and standard deviation $\sigma^{(\text{S\&P})}$ of the log-returns process $R_{t_{n,k}}^{(n,\text{S\&P})}$ of the S\&P 500 index.
The sample estimates are denoted $\hat{\mu}^{(\text{S\&P})}$ and $\hat{\sigma}^{(\text{S\&P})}$.
We apply CSYIP by assuming that the S$\&$P 500 index-return daily intensity follows (\ref{eq13_setting_CSYip}) with 
\begin{equation}
Z^{(n,\text{S\&P})}_{t_{n,k}}
	= \frac{1}{\hat{\sigma}^{(\text{S\&P})}\sqrt{\Delta t}}\left(R^{(n,\text{S\&P})}_{t_{n,k}}-\left(\hat{\mu}^{(\text{S\&P})} - \frac{{\hat \sigma}^{(\text{S\&P})^2}}{2}\right)\Delta t\right)
\label{eq_zSP}
\end{equation}
and
\begin{equation}
\xi^{(\text{S\&P})}_{t_{n,k}}
	= \sqrt{\frac{1-\mathbbm{p}^{(\text{S\&P})}_{n,\Delta t}}{\mathbbm{p}^{(\text{S\&P})}_{n,\Delta t}}}I_{\left\{Z^{(n,\text{S\&P})}_{t_{n,k}} \geq 0 \right\}}
	-  \sqrt{\frac{\mathbbm{p}^{(\text{S\&P})}_{n,\Delta t}}{1-\mathbbm{p}^{(\text{S\&P})}_{n,\Delta t}}}I_{\left\{Z^{(n,\text{S\&P})}_{t_{n,k}}   <   0 \right\}},
\label{eq_xiSP}
\end{equation}
where $\mathbbm{p}^{(\text{S\&P})}_{n,\Delta t} = \mathbb{P}\left(Z^{(n,\text{S\&P})}_{t_{n,k}} \geq 0\right) \in(0,1)$.
For simplicity we choose
$h(x)=(\sigma_h\sqrt{2\pi})^{-1}\  \exp [-x^2/(2\sigma_h^2)]$,
$g(x)=(\sigma_g\sqrt{2\pi})^{-1}\  \exp [-x^2/(2\sigma_g^2)]$,
$x\in R$ in (\ref{eq19_Rnhgs}).\footnote{
Obviously other functions in the class of CSY piecewise continuous functions are possible, including, for example, Student's $t$ or generalized hyperbolic distributions
to capture heavy tailed behavior.
The choice should be governed by the desire to obtain a ``good calibration".}
With ``$(M) = (\text{S\&P})$" in (\ref{eq19_Rnhgs}), the historical log-return time series, $R_{t_{n,k}}^{(n,\text{MSFT,hist})}$, for MSFT can be fit by the model
\begin{align}
R_{t_{n,k}}^{(n,\text{MSFT,hist})}
	&  = \nu \Delta t
	   + \sigma \sqrt{\Delta t}\xi^{(\text{S\&P})}_{n,k}
	   + \gamma \sqrt{\Delta t}\xi^{(\text{S\&P})}_{n,k} h\left(\sum_{i=1}^{k-1}\sqrt{\Delta t}\xi^{(\text{S\&P})}_{n,i}\right)
	 \nonumber \\
	&+ \delta \sqrt{\Delta t}\xi_{n,k}^{(\text{S\&P})}g\left(\sum_{j=1}^{k-1}\left(\sum_{l=1}^{j-1}\sqrt{\Delta t}\xi_{n,l}^{(\text{S\&P})}\right)\Delta t\right)
	  + \epsilon_{t_{n,k}}^{(n,\text{MSFT})},
	\label{MSFT_model}
\end{align}
where $\epsilon_{t_{n,k}}^{(n,\text{MSFT})}$ are the error terms in the above regression model denoting the MSFT stock-specific risk.
To solve for the model parameters, we construct the minimization problem
\begin{equation}
\min_{\nu>r>0,\;\sigma>0,\;\gamma>0,\;\delta>0}  \parallel \epsilon_{t_{n,k}}^{(n,\text{MSFT})} \parallel^2_2.
\label{eqof_table1}
\end{equation}

We note that the solution space of this constrained, non-linear minimization problem\footnote{
	See \cite{Coleman1996}.}
is very unstable;
small perturbations away from a parameter solution set, when used as new parameter guesses to initialize the minimization problem,
can produce a radically different solution having the same minimizing RMSE.
Table \ref{tab1_est_para} (rows S\&P${}^a$ and S\&P${}^b$) present the parameter estimates for two solutions which,  to three significant digits, have the same values
for RMSE.
Of the two solutions, S\&P${}^a$ is more realistic since $\sigma = 0$ for S\&P${}^b$.
\begin{table}[h]
\caption{Parameter estimates obtained from the minimization problem (\ref{eqof_table1}).}
\label{tab1_est_para}
    \begin{center}		
	\begin{tabular}{c c c c c c c c}
		\toprule
		factor & $\nu$ & $\sigma$ & $\gamma$ & $\sigma_h$ & $\delta$ & $\sigma_g$ & RMSE \\
		\midrule
		S\&P${}^a$ & 0.0016 & 0.0020 & 0.29 & 8.8 & 0.089 & $1.8 \cdot 10^3$ & 0.0214 \\
		S\&P${}^b$ & 0.0016 & $1.1 \cdot 10^{-10}$ & 0.37 & 9.9 & 0.44 & 0.052 & 0.0214  \\
		S\&P Hist & $2.2\cdot 10^{-14}$ & 0.014 & 0.0014 & 0.13 & 1.09 & 0.10 & 0.0214 \\
		& & & & & & & \strut \\
		\multicolumn{8}{l}{S\&P with ARMA(1,1)-GJR GARCH(1,1) innovations modeled by:}\\
		ghyp    & 0.0013 & $2.2\cdot 10^{-14}$ & 0.0078 & 0.073 & 0.35 & 0.072 & 0.0254 \\
		stud $t$ & 0.0017 & $4.4\cdot 10^{-14}$ & 0.011 & 0.22 & 1.1 & 0.10 & 0.0252 \\
		norm   & 0.0018 & $2.2\cdot 10^{-14}$ & 0.0038 & 0.077 & 1.1 & 0.10 & 0.0253 \\
		& & & & & & & \strut \\
		J${}^a$      & 0.0019 & 0.0065 & 0.083 & 73 & 0.11 & $7.6\cdot 10^4$ & 0.0244 \\
		J${}^b$      & 0.0020 & 0.0062 & 0.0000 & 0.0027 & 0.84 & $1.1\cdot 10^2$ & 0.0244 \\
		FF3${}^a$  & 0.0019 & 0.0049 & 0.012 & $9.7 \cdot 10^3$ & 0.0008 & $1.0\cdot 10^4$ & 0.0249 \\
		FF3${}^b$  & 0.0018 & 0.0044 & 0.0068 & 0.59 & 1.09 & $3.1\cdot 10^5$ & 0.0249 \\
		\bottomrule
	\end{tabular}
    \end{center}
\end{table}

Equation (\ref{MSFT_model}) models MSFT returns in terms of systematic risk from the S\&P 500 index based on the
up- and down-turns of the centralized log-return (\ref{eq_zSP}).
We consider the case where the systematic risk is determined directly by the up- and down-turns of $R^{(n,\text{S\&P})}_{t_{n,k}}$, i.e.
\begin{equation*}
	\xi_{n,k}^{(\text{S\&P})} = 
	\begin{cases}
	 \ \ 1 \quad \textup{if} \quad R^{(n,\text{S\&P})}_{t_{n,k}} \ge 0,
	\\
	-1 \quad \textup{if} \quad R^{(n,\text{S\&P})}_{t_{n,k}} < 0.
	\end{cases}
\end{equation*}
This has the effect of setting $\mathbbm{p}_{n,\Delta t}^{(\text{S\&P})} = p_{n,\Delta t}^{(\text{S\&P})}$.
One solution of this subsequent minimization (\ref{eqof_table1}) are shown in Table \ref{tab1_est_para} (row S\&P Hist).
In terms of RMSE this approach appears equivalent to the first.

We have also explored a third option for the minimization (\ref{MSFT_model}).\footnote{
	It is tempting to include the historical data on upturns of the log-return series of MSFT  as its own ``index".
	However, this is inherently inconsistent with our findings that the stock price dynamics does not follow the
	discrete dynamics of a generalized Brownian motion.}
The centralized return $Z_{n,k}^{(M)}$ defined in (\ref{eq13_setting_CSYip}) implicitly treats $R_{t_{n,k}}^{(n,M)}$ as if they were i.i.d. normally distributed (Gaussian noise).
Knowing this is not true,
we performed an ``ARMA(1,1) - GJR GARCH(1,1) with assumed innovation distribution" fit
\begin{align}
    \begin{split}
	R_{k,\Delta t}^{(n,\text{S\&P})} & = \mu + \phi_1 ( R_{k-1,\Delta t}^{(n,\text{S\&P})} - \mu ) + a_k + \theta_1 a_{k-1}, \\
	a_k & = \sigma_k \epsilon_k, \\
	\sigma_k^2 & = \alpha_0 + ( \alpha_1 + \gamma_1 I_{k-1} ) a_{k-1}^2 + \beta_1 \sigma_{k-1}^2,
    \end{split}
    \label{eqn_AG}
\end{align}
 to the log-returns $R_{k,\Delta t}^{(n,\text{S\&P})}$ of the market index.
For the innovation distribution we assumed either standardized-Gaussian, -Student's $t$, or -generalized hyperbolic.
Denote the resultant time series of residuals obtained from this ARMA - GJR GARCH fit as $\epsilon_{n,k}^{AG,D}$ where $D$ identifies the distribution used in the fit.
By filtering out the ARMA-GARCH factors, the series $\epsilon_{n,k}^{AG,D}$ should be "noise" ( though perhaps not Gaussian).
Following (\ref{eq13_setting_CSYip}), the time series $\epsilon_{n,k}^{AG,D}$ replaces $R_{k,\Delta t}^{(n,\text{S\&P})}$ in (\ref{eq_zSP}) and the subsequent steps leading
to the construction of the minimization problem (\ref{MSFT_model}).

\begin{table}[t]
\caption{ARMA(1,1)-GJR GARCH(1,1) with standardized Student's $t$ and generalized hyperbolic innovation fits to the residuals $\epsilon_{t_{n,k}}^{(n,\text{MSFT})}$ of (\ref{MSFT_model}).}
	\label{tab_AG}
    \begin{center}		
	\begin{tabular}{c  c c c c  c c c c}
		\toprule
		          & \multicolumn{4}{c}{Student's $t$} & \multicolumn{4}{c}{Generalized hyperbolic} \\
		          \cmidrule(r){2-5}\cmidrule(r){6-9}
		parameter & estimate & SE & $t$-value & $p$-value & estimate & SE & $t$-value & $p$-value \\
		$\mu$      & $4.0 \cdot 10^{-5}$ & $6\cdot 10^{-4}$ & 0.067   & 0.94 &  $4.0 \cdot 10^{-5}$ & $6.5\cdot 10^{-4}$ & 0.06 & 0.95 \\
		$\phi_1$    & $0.32$  & 0.37       & 0.87     & 0.38     & 0.30     & 0.40 & 0.77      & 0.44 \\
		$\theta_1$ & $-$0.45 & 0.35       & $-$1.3  & 0.19     & $-$0.44 & 0.37 & $-$1.2   & 0.24 \\
		$\alpha_0$ & $1.2 \cdot 10^{-5}$ & $3.7 \cdot 10^{-5}$ & 0.32 & 0.75 & $1.2 \cdot 10^{-5}$ & $9.0 \cdot 10^{-6}$ & 1.46 & 0.14 \\
		$\alpha_1$ & 0.24     & 0.35       & 0.68     & 0.50     & 0.24     & 0.13  & 1.82     & 0.07\\
		$\beta_1$  & 0.727   & 0.024      & 29       & 0          & 0.724   & 0.05  & 14        & 0 \\
		$\gamma_1$ & $-$0.10 & 0.14    & $-$0.67 & 0.51     & $-$0.11 & 0.16 & $-$0.68 & 0.50 \\
		shape ($\nu^a,\zeta^b$) & 13        & 6.3          & 2.05     & 0.04     & 2.5       & 58    & 0.04     & 0.97 \\
		skew ($\rho^b$) &             &               &           &            & $-$0.31 & 7.5   & $-$0.04 & 0.97 \\
		$\lambda^b$ &        &               &           &            & $-$5.8   & 42    & $-$0.14 & 0.89 \\\
		 & & & & & & & & \\
		\multicolumn{9}{l}{Robust standard errors} \\
		parameter & estimate & SE & $t$-value & $p$-value & estimate & SE & $t$-value & $p$-value \\
		$\mu$      & $4.0 \cdot 10^{-5}$ & 0.0024 & 0.017   & 0.99 &  $4.0 \cdot 10^{-5}$ & $7.4\cdot 10^{-4}$ & 0.054 & 0.96 \\
		$\phi_1$    & $0.32$  & 0.79       & 0.41     & 0.68     & 0.30     & 0.51 & 0.60      & 0.55 \\
		$\theta_1$ & $-$0.45 & 0.64       & $-$0.71  & 0.48     & $-$0.44 & 0.44 & $-$1.0   & 0.32 \\
		$\alpha_0$ & $1.2 \cdot 10^{-5}$ & $3.6 \cdot 10^{-4}$ & 0.034 & 0.97 & $1.2 \cdot 10^{-5}$ & $2.6 \cdot 10^{-5}$ & 0.48 & 0.63 \\
		$\alpha_1$ & 0.24     & 3.4         & 0.070     & 0.94     & 0.24     & 0.38  & 0.64     & 0.52\\
		$\beta_1$  & 0.727   & 0.52      &1.4       & 0.16          & 0.724   & 0.089  & 8.1    & 0 \\
		$\gamma_1$ & $-$0.10 & 0.18    & $-$0.54 & 0.59     & $-$0.11 & 0.31 & $-$0.34 & 0.73 \\
		shape ($\nu^a,\zeta^b$) & 13        & 110         & 0.12     & 0.91     & 2.5       & 343    & 0.007     & 0.99 \\
		skew  ($\rho^b$) &             &               &           &            & $-$0.31 & 43   & $-$0.007 & 0.99 \\
		$\lambda^b$ &        &               &           &            & $-$5.8   & 229    & $-$0.026 & 0.98 \\
		\bottomrule
		\multicolumn{9}{l}{${}^a$\small{$\nu$ is the degrees of freedom for the standardized Student's $t$ distribution.}}\\
		\multicolumn{9}{l}{${}^b$\small{The rugarch package in {\bf R} uses the $\zeta, \rho, \lambda$ parametrization for the
		 standardized generalized hyperbolic distribution.}}\\
		\multicolumn{9}{l}{\ \small{ See {\it Introduction to the rugarch package. (Version 1.4-3)} A. Ghalanos, July 15, 2020.}} \\
	\end{tabular}
    \end{center}
\end{table}
The results of the ARMA - GJR GARCH fit using the Student's $t$ and generalized hyperbolic distributions are given in Table \ref{tab_AG}.
Parameter solutions for the minimization problem (\ref{MSFT_model}) using the residuals $\epsilon_{n,k}^{AG,norm}$, $\epsilon_{n,k}^{AG,stud\; t}$,
and $\epsilon_{n,k}^{AG,ghyp}$ are shown in the correspondingly labeled rows in Table \ref{tab1_est_para}.
Based upon RMSE values, this option provides less predictive power than direct use of $R_{k,\Delta t}^{(n,\text{S\&P})}$.

In addition to using the log-return process of the S\&P 500 index as a model for the MSFT log-return dynamics,
we also considered models based upon time series of “implied alphas”.
The first is based upon Jensen's alpha\footnote{
	See \cite{Jensen1968} and \cite{Jensen1969}; \cite{Aragon2006}; \cite{Breloer2016}.}
while the second is the implied alpha determined from the Fama-French three-factor (FF3) model\footnote{
	See \cite{Fama1993}.
	More general factor models can also be considered (e.g.\cite{Carhart1997}, \cite{Fama2004}, \cite{Fama2015}, and \cite{Fama2017}).}.

The Jensen alpha time-series was computed as
\begin{equation}
\alpha_{k\Delta t}^{(\text{J})} = R^{(\text{MSFT})}_{k\Delta t} - R_{f,k\Delta t}
		   - \beta^{(\text{J})} \left [ R^{(\text{S\&P})}_{k\Delta t} - R_{f,k\Delta t} \right ]
\label{eq_jensen}
\end{equation}
where $k\Delta t$ ran over the daily trading days from 07/01/2019 through 06/30/2020.
Daily return values for the risk-free rate $R_{f,k\Delta t}$ were computed from the
10-year Treasury rates and returns from the S\&P 500 were used to represent the market.
Single values for $\beta^{(\text{J})}$ and $\alpha^{(\text{J})}$ were computed using a linear regression for
(\ref{eq_jensen}) over the entire data set.
A time series of daily values of $\alpha_{k\Delta t}^{(\text{J})}$ were then computed using (\ref{eq_jensen}).
The values of this implied Jensen's alpha time-series replaced the S\&P 500 index values in the above analysis (\ref{eq_xiSP}) through (\ref{eqof_table1}).

A time series of implied values of $\alpha_{k\Delta t}^{(\text{FF3})}$ were computed analogously.
Daily values for $R_m$, $R_f$, $R_{\text{SMB}}$ and $R_{\text{HML}}$ were obtained from the U.S. Research
Returns Data page of Professor French's web site\footnote{
	See {\it https://mba.tuck.dartmouth.edu/pages/faculty/ken.french/data$\_$library.html$\#$Research.}
For explanation of the data see {\it https://mba.tuck.dartmouth.edu/pages/faculty/ken.french/Data$\_$Library/f-f$\_$factors.html.}}.
Again, single values for the FF3 constants $\beta^{(M)}$, $\beta^{(\text{SMB})}$ and $\beta^{(\text{HML})}$
were obtained from linear regression using the entire data set.
Using these, daily values for $\alpha_{k\Delta t}^{(\text{FF3})}$ were computed.

Parameter values obtained from the minimization problem (\ref{eqof_table1}) for each implied alpha time series are also displayed in
Table \ref{tab1_est_para} (rows J and FF3 respectively).
Based on RMSE values, the Jensen and FF3 alpha models are poorer predictors than $R_{k,\Delta t}^{(n,\text{S\&P})}$ for the dynamics of the
returns of MSFT within the discrete pricing model (\ref{eq19_Rnhgs}), but are better than option 3.

To test for clustering of the volatility and/or heavy tails, we fit the residuals $\epsilon_{t_{n,k}}^{(n,\text{MSFT})}$ in (\ref{MSFT_model}) to an
ARMA(1,1)-GARCH(1,1) model (equation (\ref{eqn_AG}) with $\gamma_1 = 0$) with Student's $t$ innovations.
Using the residuals  $\epsilon_{t_{n,k}}^{(n,\text{MSFT})}$ obtained from the fit in row S\&P${}^a$ in Table \ref{tab1_est_para},
Table \ref{tab2_est_para} reports the results for the fitted parameters as well as the $p$-values obtained.
The {\it p}-values indicate that the fit to the GARCH component of the model (\ref{eqn_AG}) is significant at the 5\% level,
while the fits to the Student's {\it t}-distribution (degrees of freedom) and, particularly,
the ARMA component are much less so. 
\begin{table}[h]
\caption{Parameters estimates for fitting the sample residuals $\epsilon_{t_{n,k}}^{(n,MSFT)}$ to the ARMA-GARCH form of (\ref{eqn_AG}).}
			\label{tab2_est_para}
	\begin{center}		
			\begin{tabular}{c c c c c}
				\toprule
				\textrm{}&{Estimate}&{Standard Error}&{{\it t}-statistic}&{{\it p}-value} \\
				\midrule		
				$\mu$       & 0.00 & 0.00 &  0.52  & 0.6 \\
				$\phi_1$    & 0.33     & 0.44 & 0.74      & 0.45 \\
				$\theta_1$ & $-$0.45 & 0.42 & $-$1.08 & 0.28 \\
				$\alpha_0$ & 0.00     & 0.00 & 2.08     & 0.04 \\
				$\alpha_1$ & 0.71     & 0.07 & 10.85   & 0.00 \\
				$\beta_1$ & 0.25     & 0.08 & 3.36     & 0.00 \\
				$\nu^a$   & 13.7     & 12.75 & 1.07     & 0.28 \\
				\bottomrule
				\multicolumn{5}{l}{${}^a\nu$ is degrees of freedom of Student's $t$ distribution.} \\
			\end{tabular}
	\end{center}
\end{table}

\begin{figure}
 	\begin{center} 
    	\centering\includegraphics[width=0.4\textwidth]{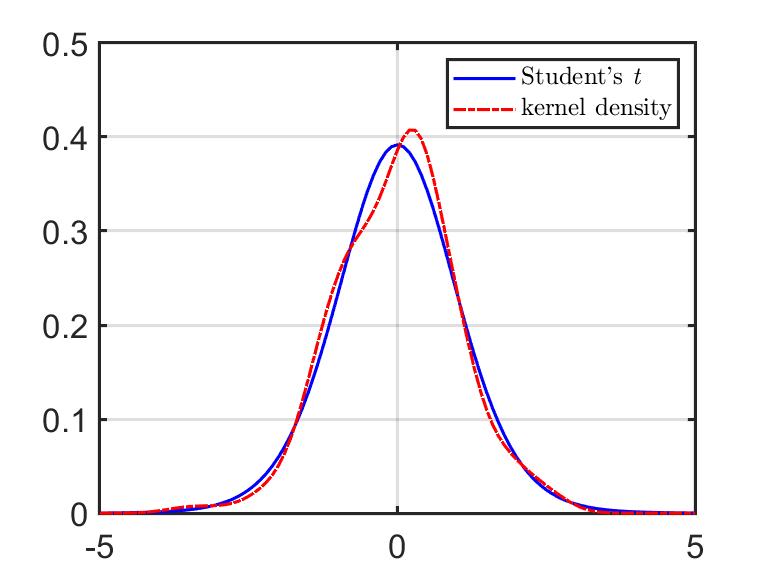}
    	\caption{Comparison of the standardized Student's {\it t}-distribution with $\nu=13$ compared to the empirical distribution
			of $\epsilon_k$ computed from (\ref{eqn_AG}).}
    	\label{fig7_pdf}
    \end{center}
\end{figure}

Figure \ref{fig7_pdf} compares the fitted Student's {\it t}-distribution for 13 degrees of freedom with the empirical values of $\epsilon_k$
computed from the ARMA-GARCH form of (\ref{eqn_AG}) using the parameters in Table \ref{tab2_est_para}.
The comparison shows that the tails of the distribution of sample residuals, $\epsilon_{t_{n,k}}^{\ast}$, are not exponentially bounded.
In addition, the sample residual distribution is not symmetric, but is skewed to the right.
Overall, the model (\ref{eqn_AG}) does not capture the empirical phenomena displayed by the MSFT stock over this time period.

\section{Option Pricing when the Underlying Stock Log-returns are Path-dependent}
\label{sec5}

\noindent
In this section, we use the path-dependent dynamics of the underlying stock price in continuous-time as given by (\ref{eq18_Sconti}), and in discrete-time as given by (\ref{eq17_Snhgs}),
to derive the corresponding risk-neutral dynamics and to value the ECC $\mathcal{C}$.
We first consider the continuous-time stock price dynamics $S_t^{(h,g,S)},\;t\in[0,T]$ on  $\left(\Omega,\mathbb{F} = \left\{\mathcal{F}_t = \sigma(B_u)\right\}_{u \leq t},\mathbb{P}\right)$ in (\ref{eq18_Sconti}).
Using It\^{o}'s formula, we have
\begin{align}
	dS_t^{(h,g,S)} &= N_1 S_t^{(h,g,S)}dt + N_2 S_t^{(h,g,S)}dB_t,
	\label{eq20_dSt_P}
\end{align}
where
\begin{align*}
	N_1 &= \nu +\frac{1}{2}\sigma^2+\frac{1}{2}\gamma^2{h(B_t)}^2+\frac{1}{2}\delta^2\left(g\left(\int_0^{t} B_{u}du\right) \right)^2 + \sigma \gamma h(B_t) +\sigma \delta g\left(\int_0^{t} B_{u}du\right) \\
       & \qquad + \gamma \delta h(B_t)  g\left(\int_0^{t} B_{u}du\right),\\
	N_2 &= \sigma+\gamma h(B_t)+\delta g\left(\int_0^{t} B_{u}du \right).
\end{align*}
From (\ref{eq20_dSt_P}), the risk-neutral stock price dynamics are defined by 
\begin{align}
	dS_t^{(h,g,S)} &= r S_t^{(h,g,S)}dt + N_2 S_t^{(h,g,S)}dB_t^{(\mathbb{Q})},\; t\in[0,T],
	\label{eq21_dSt_Q}
\end{align}
where $S_0^{(h,g,S)}=S_0^{(S)}$.
In (\ref{eq21_dSt_Q}), $B_t^{\mathbb{Q}} = B_t+\int ^t_0 \theta_s ds$ is a BM on $\mathbb{Q}$
and the market price of risk $\theta_t,\;t\in [0,T]$, is given by $\theta_t = (N_1-r)/N_2$.
Because $\theta_t$ is uniquely determined for $t \ge 0$, $(\mathcal{S},\mathcal{B})$ is an arbitrage-free and complete market.
Let $f_t=f(S_t,t),t\in[0,T]$, denote the price dynamics of the ECC $\mathcal{C}$ with terminal time $T>0$
and final payoff $f_T=G(S_T)$.
Then, $f_0=e^{-rT} \mathbb{E}^{(\mathbb{Q})} G(S_T)$ under the usual regularity conditions \citep{Duffie2001}.

Consider the discrete filtration $\mathbb{F}^{(d)} = \left\{\mathcal{F}_0 = \{\varnothing,\Omega\},\mathcal{F}_k = \sigma(\xi_{n,1}^{(M)},\ldots,\xi_{n,k}^{(M)}),k = 1,\ldots,n,\;n \in \mathcal{N}\right\}$, where $\xi_{n,k}^{(M)}$ is given by (\ref{eq13_setting_CSYip}).
Let
\begin{equation}
	\eta^{(M)}_{k,\Delta t} = \sigma+ \gamma h\left(\sum^{k}_{i = 1}\sqrt{\Delta t} \xi_{n,i+1}^{(M)} \right)+ \delta g \left(\sum^{k}_{j = 1}\left(\sum^{j-1}_{l = 1} \sqrt{\Delta t}\xi_{n,l}^{(M)} \right)\Delta t\right)>0,\;k=1,\ldots,n,
	\label{eq22_eta}
\end{equation}
with $\eta^{(M)}_{0,\Delta t} = \sigma,\;\nu>r>0,\;\sigma>0,\;\gamma>0,\;\delta>0$,
and $h(x)\geq 0,\;g(x)\geq 0,\;x\in R$ are uniformly bounded piecewise continuous functions.
According to (\ref{eq19_Rnhgs}), the stock binomial tree dynamics conditioned on $\mathcal{F}_{k}$ is given by
\begin{align}
	S_{(k+1)\Delta t}^{(n,h,g,S)} &= S_{k\Delta t}^{(n,h,g,S)} \exp \left\{\nu \Delta t+\eta^{(M)}_{k,\Delta t}\sqrt{\Delta t}\xi_{n,k+1}^{(M)}\right\},\;S_0^{(h,g,S)} =S_0,\;k=0,1,\ldots,n-1.
	\label{eq23_S_discrete}
\end{align}
Note that $\eta^{(M)}_{k,\Delta t}$ represents the time-varying stock volatility at $k\Delta t$ as a function of index intensities $\xi_{n,i}^{(M)},\;i=1,\ldots,n$.
From (\ref{eq13_setting_CSYip}), conditionally on $\mathcal{F}_{k},\;k=0,\ldots,n-1$, we have
\begin{equation}
	S_{(k+1)\Delta t}^{(n,h,g,S)}=
	\begin{cases}
	S_{(k+1)\Delta t}^{(n,h,g,S,u)}&=S_{k\Delta t}^{(n,h,g,S)}\exp\left\{\nu \Delta t+\eta^{(M)}_{k,\Delta t}\sqrt{\Delta t}\sqrt{\frac{1-\mathbbm{p}^{(M)}_{n,\Delta t}}{\mathbbm{p}^{(M)}_{n,\Delta t}}} \right\},\;\textup{w.p.}\;\mathbbm{p}^{(M)}_{n,\Delta t},
	\\
	S_{(k+1)\Delta t}^{(n,h,g,S,d)}&=S_{k\Delta t}^{(n,h,g,S)}\exp\left\{\nu \Delta t-\eta^{(M)}_{k,\Delta t}\sqrt{\Delta t}\sqrt{\frac{\mathbbm{p}^{(M)}_{n,\Delta t}}{1-\mathbbm{p}^{(M)}_{n,\Delta t}}}\right\},\;\textup{w.p.}\;1-\mathbbm{p}^{(M)}_{n,\Delta t}.
	\end{cases}
	\label{eq24_S}
\end{equation}
Then, similarly to the derivation of (\ref{eq8_qp}), we obtain the conditional risk-neutral probabilities
\begin{equation}
	\mathbbm{q}_{k,\Delta t}^{(S)}
		= \mathbbm{p}_{n,\Delta t}^{(M)} - \theta^{(S)}_{k,\Delta t}\sqrt{\mathbbm{p}_{n,\Delta t}^{(M)}(1-\mathbbm{p}_{n,\Delta t}^{(M)})\Delta t},
		   \quad \theta^{(S)}_{k,\Delta t} = \frac{\nu-r}{\eta^{(M)}_{k,\Delta t}},\; k=0,\ldots,n-1,\;n\in\mathcal{N}.
	\label{eq25_qp}
\end{equation}
Conditionally on $\mathcal{F}_{k}$, the stock risk-neutral dynamics is given by $S_{0}^{(n,\mathbbm{q})} = S_0$, and for $k = 0,\ldots,n-1$,
\begin{equation}
	S_{(k+1)\Delta t}^{(n,\mathbbm{q})}=
	\begin{cases}
	S_{(k+1)\Delta t}^{(n,\mathbbm{q},u)}&=S_{k\Delta t}^{(n,\mathbbm{q})}\exp\left\{\nu \Delta t+\eta^{(M)}_{k,\Delta t}\sqrt{\Delta t}\sqrt{\frac{1-\mathbbm{p}^{(M)}_{n,\Delta t}}{\mathbbm{p}^{(M)}_{n,\Delta t}}} \right\},\;\textup{w.p.}\;\mathbbm{q}^{(S)}_{k,\Delta t},
	\\
	S_{(k+1)\Delta t}^{(n,\mathbbm{q},d)}&=S_{k\Delta t}^{(n,\mathbbm{q})}\exp\left\{\nu \Delta t-\eta^{(M)}_{k,\Delta t}\sqrt{\Delta t}\sqrt{\frac{\mathbbm{p}^{(M)}_{n,\Delta t}}{1-\mathbbm{p}^{(M)}_{n,\Delta t}}}\right\},\;\textup{w.p.}\;1-\mathbbm{q}^{(S)}_{k,\Delta t}.
	\end{cases}
	\label{eq26_Sq}
\end{equation}
In (\ref{eq25_qp}), the market price of risk $\theta^{(S)}_{k,\Delta t}$ is time dependent, leading to a heavy-tailed distribution for the log-returns $R_{(k+1)\Delta t}^{(n,\mathbbm{q})} = \ln \left(
S_{(k+1)\Delta t}^{(n,\mathbbm{q})}/S_{k\Delta t}^{(n,\mathbbm{q})}\right)$ in (\ref{eq26_Sq}), which can produce a more realistic pricing model for the EEC $\mathcal{C}$. The risk-neutral price $f_{k \Delta t}^{(n)} =f\left( S_{k \Delta t}^{(n,\mathbbm{q})},k\Delta t\right)$ of $\mathcal{C}$ is given by
\begin{equation*}
	f_{k \Delta t}^{(n)} = e^{-r\Delta t}\left\{\mathbbm{q}^{(S)}_{k,\Delta t}f_{k \Delta t}^{(n,u)}+(1-\mathbbm{q}^{(S)}_{k,\Delta t})f_{k \Delta t}^{(n,d)}\right\},
\end{equation*}
where $f_{(k+1) \Delta t}^{(n,u)} =f\left( S_{(k+1) \Delta t}^{(n,\mathbbm{q},u)},(k+1)\Delta t\right)$, $f_{(k+1) \Delta t}^{(n,d)} =f\left( S_{(k+1) \Delta t}^{(n,\mathbbm{q},d)},(k+1)\Delta t\right)$, $k = 0,\ldots,n-1$, and $f_{n \Delta t}^{(n)} = f_{T}^{(n)} = G(S_T)$.

\subsection{An Example}

\begin{figure}
 	\begin{center} 
    	\centering\includegraphics[width=0.5\textwidth]{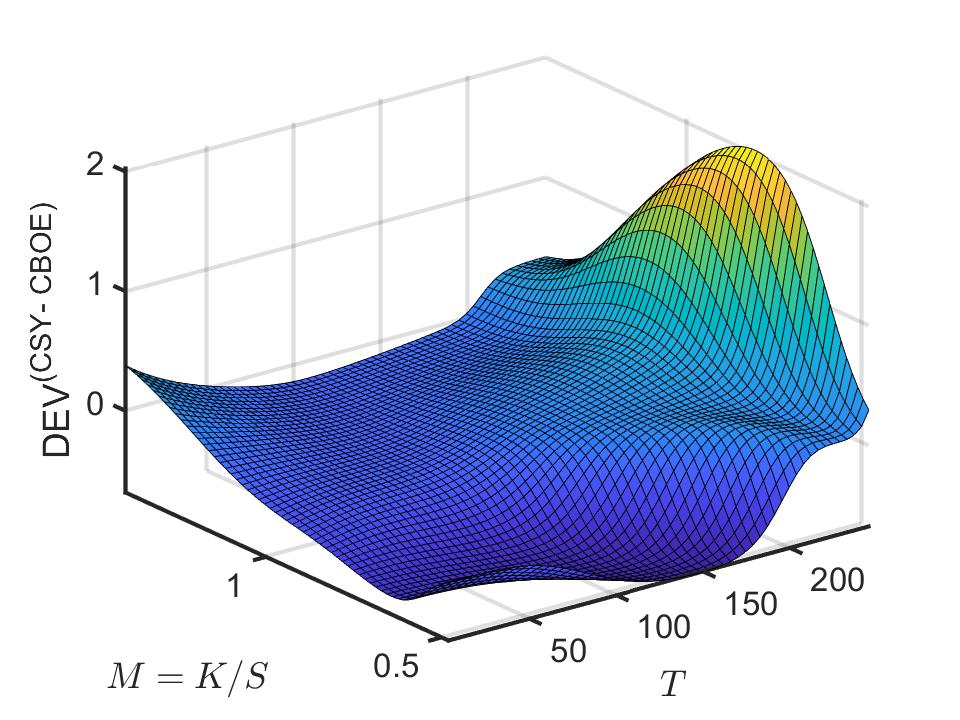}
    	\caption{Relative volatility deviation against time to maturity $T$ and moneyness $M=K/S$.}
    	\label{fig_dev}
    \end{center}
\end{figure}

\noindent
We apply CSYIP to generate an artificial binomial tree to obtain call option prices using the daily closing prices and the corresponding log-return process
of MSFT combined with the parameter estimations of row S\&P${}^a$ in Table \ref{tab1_est_para}.
Using the daily intensity dynamics $\xi^{(\text{S\&P})}_{t_{n,k}}$ in (\ref{eq_xiSP}), we have
\begin{equation}
	\eta^{(\text{S\&P})}_{k,\Delta t} = \hat{\sigma}+ \hat{\gamma} h\left(\sum^{k}_{i = 1}\sqrt{\Delta t} \xi_{n,i+1}^{(\text{S\&P})} \right)+ \hat{\delta} g \left(\sum^{k}_{j = 1}\left(\sum^{j-1}_{l = 1} \sqrt{\Delta t}\xi_{n,l}^{(\text{S\&P})} \right)\Delta t\right),
	\label{eq_eta53}
\end{equation}
where $\hat{\sigma}, \hat{\gamma}$, and $\hat{\delta}$ are given in row S\&P${}^a$ of Table \ref{tab1_est_para}.
Following the framework developed in (\ref{eq20_dSt_P}) through (\ref{eq26_Sq}) based on the MSFT price data and the 10-year Treasury rate\footnote{
We set $07/01/2019$ as $t=0$. Based on the information on $07/01/2019$, the initial capital $S_0^{(\text{MSFT})} = 135.68$,
and annual risk-free rate $r = 2.03\%$.},
we obtain the call option price $C^{(\text{MSFT;CSY})}_t ( S^{(\text{MSFT})}_{0},\linebreak[1] K,\linebreak[1] T,\linebreak[1] \hat{\nu},\linebreak[1] \hat{\sigma},\linebreak[1] \hat{\gamma},\linebreak[1] \hat{\delta} )$.
Using the Black-Scholes-Merton formula\footnote{See \cite{Hull2018}, Chapter 15.},
we calculate the corresponding CSY implied volatility $\sigma^{(\text{CSY}-\text{vol})}$.
We compare it to the implied volatility based on the call options data for MSFT from the Chicago Board Option Exchange (CBOE),
which we denote as $\sigma^{(\text{CBOE}-\text{vol})}$, by defining the relative volatility deviation,
\begin{equation*}
\textrm{DEV}^{(\text{CSY}-\text{CBOE})} = \frac{\sigma^{(\text{CSY}-\text{vol})}-\sigma^{(\text{CBOE}-\text{vol})}}{\sigma^{(\text{CSY}-\text{vol})}}.
\end{equation*}
$\textrm{DEV}^{(\text{CSY}-\text{CBOE})}$ can be used to identify mispricing of various option contracts based on the deviation of the existing market implied volatility surface from the theoretical $\sigma^{(\text{CSY}-\text{vol})}$.
Figure \ref{fig_dev} shows values of DEV$^{(\text{CSY}-\text{CBOE})}$ ranging from $-0.68$ to $2.01$, with an increasing trend as $T$ increases.

\section{Option Pricing for Markets with Informed Traders}
\label{sec6}

\noindent
We extend the results of Section \ref{sec5} to option pricing for markets with informed traders.
Suppose that at $(k-1)\Delta t,\;k=1,\ldots,n,\;n\Delta t = T$, a trader (denoted by $\aleph$) observed the market index closing values $S_{j\Delta t}^{(n,M)},\;j=0,1,\ldots,k-1$,
with the corresponding values of $Z^{(n,M)}_{t_{n,k}}$ given by (\ref{eq13_setting_CSYip}(a)).
Consequently, $\aleph$ knows the probability $\mathbbm{p}_{\Delta t}^{(\aleph,M)}\in (1/2,1)$ for the sign of $Z_{k\Delta t}^{(n,M)}.$\footnote{
By observing the price trajectory $S_{\Delta t}^{(n,M)},\;j=0,1,\ldots,k-1$, $\aleph$ knows whether the log-return process
$R_{k\Delta t}^{(n,M)}= \ln\left(S_{k\Delta t}^{(n,M)}/S_{(k-1)\Delta t}^{(n,M)}\right)$ is above or below the “benchmark level” 
$\left(\mu^{(M)}-{\sigma^{(M)}}^2/2\right)\Delta t$ with probability $p_{\Delta t}^{(\aleph,M)}$.}
$\aleph$ assumes that the continuous time dynamics of the stock price $S_t^{(h,g,S)}$ is given by (\ref{eq18_Sconti}).
The risk-free asset dynamics is given by
\begin{equation}
	\beta_t = \beta_0 e^{\int^t_0 r_u du},\; \beta_0>0,\;t \in [0,T],
	\label{eq28_bond}
\end{equation}
where $r_t,\;t\in[0,T]$ is the instantaneous risk-free rate.\footnote{We assume that $r_t$ has a continuous first derivative.} 

The discrete filtration $\mathbb{F}^{(d)}$ defined in Section \ref{sec5} defines the discrete dynamics of the market index.
From (\ref{eq24_S}), define the binomial tree conditioned on $\mathcal{F}_k$ as
\begin{equation}
	S_{(k+1)\Delta t}^{(n,h,g)}=
	\begin{cases}
	S_{(k+1)\Delta t}^{(n,h,g,u)}&=S_{k\Delta t}^{(n,h,g)}\exp\left\{U_{n,k+1}^{(h,g)}\right\},\;\textup{w.p.}\;\mathbbm{p}^{(M)}_{n,\Delta t},
	\\
	S_{(k+1)\Delta t}^{(n,h,g,d)}&=S_{k\Delta t}^{(n,h,g)}\exp\left\{D_{n,k+1}^{(h,g)}\right\},\;\textup{w.p.}\;1-\mathbbm{p}^{(M)}_{n,\Delta t},
	\end{cases}
	\label{eq29_S}
\end{equation}
for $k = 0, \ldots, n-1$ with $S_0^{(n, h, g)} = S_0$. 
In (\ref{eq29_S}), with the drift terms $\nu$ not assumed time dependent $\nu_{k\Delta t}$, we have:
\begin{itemize}
	\item[(i)] $U_{n,k+1}^{(h,g)} = \nu_{k\Delta t} \Delta t+\eta^{(M)}_{k,\Delta t}\sqrt{\Delta t}\sqrt{\frac{1-\mathbbm{p}^{(M)}_{n,\Delta t}}{\mathbbm{p}^{(M)}_{n,\Delta t}}}$
	and $D_{n,k+1}^{(h,g)} = \nu_{k\Delta t} \Delta t-\eta^{(M)}_{k,\Delta t}\sqrt{\Delta t}\sqrt{\frac{\mathbbm{p}^{(M)}_{n,\Delta t}}{1-\mathbbm{p}^{(M)}_{n,\Delta t}}}$;
	\item[(ii)] $\eta^{(M)}_{k,\Delta t}$ is defined in (\ref{eq22_eta}) with $\sigma > 0$,\;$\gamma > 0$, and $h(x)$ and $g(x)$ are CSY piecewise continuous functions;
	\item[(iii)] $\nu_{k\Delta t}-\frac{1}{2} {\eta^{(M)}_{k,\Delta t}}^2 > r_{k\Delta t} > 0$,\; which requires $h(x)\geq 0$ and $g(x)\geq 0$;\footnote{
		To have the inequalities, $\nu_{k\Delta t}-\frac{1}{2}{\eta^{(M)}_{k,\Delta t}}^2 > r_{k\Delta t} > 0,\;k=0,\ldots,n$, satisfied, we assume functions $h(x)\geq 0$ and $g(x)\geq 0,\;x \in R$ are uniformly bounded,
		and furthermore, $\min_{t\in[0,T]}\left(\mu_t - r_t\right)>0$ is sufficiently large.}
	\item[(iv)] $\mathbbm{p}^{(M)}_{n,\Delta t}=\mathbb{P}\left(Z_{k\Delta t}^{(n,M)}> 0\right)=\mathbb{P}\left(\xi_{n,k}^{(M)}> 0\right)$,
		assuming that  $\mathbb{P}\left(Z_{k\Delta t}^{(n,M)}= 0\right)=\mathbb{P}\left(\xi_{n,k}^{(M)}= 0\right) = 0$.
\end{itemize}
With the conditions $\mathbbm{p}_{n,\Delta t}^{(M)} = p^{(0)} + p^{(1)} \sqrt{\Delta t} + p^{(2)} \Delta t$ where $p^{(0)} \in (0,1)$, $p^{(1)} \in R$, $p^{(2)} \in R$ and $\Delta t = O(T/n)$,
then the tree (\ref{eq29_S}) is recombining.

\subsection{Forward contract strategy}
\label{sec61}

\noindent
At time $k\Delta t,k=0,\ldots,n-1$, $\aleph$ places binary bets on whether $\xi_{n,k+1}^{(M)}\geq 0$ or $\xi_{n,k+1}^{(M)}<0$.
The outcome $o^{(M)}_{n,k+1}$ of the bet is $\mathcal{F}_k$-measurable,
with (a) $o^{(M)}_{n,k+1}=1$ if $\aleph$ has guessed the sign of $\xi_{n,k+1}^{(M)}$ correctly;
otherwise (b) $o^{(M)}_{n,k+1}=0$.
We assume that $\mathbb{P}(o^{(M)}_{n,k+1} = 1) = 1-\mathbb{P}(o^{(M)}_{n,k+1} = 0) = p^{(\aleph,M)}_{\Delta t}$.
Let $\Delta _{ k\Delta t }^{ (\aleph, M) } =N _{ k\Delta t }^{ (\aleph, M) } / S_{ k\Delta t }^{ (n,h,g) }$,
where an optimal value for $N _{ k\Delta t }^{ (\aleph, M) }$ is determined below.\footnote{
The parameter $N_{k\Delta t}^{(\aleph,M)}$ will be optimized and will enter the formula for the non-negative yield that $\aleph$ will receive when
trading options; see Section \ref{sec62}.
And the long position in the forward contract could be taken by any trader who believes that
$S_{(k+1)\Delta t}^{(n,h,g)} = S_{(k+1)\Delta t}^{(n,h,g,u)}$ is more likely to happen.}
If at $k \Delta t$, $\aleph$ believes that the event $\left\{\xi_{n,k+1}^{(M)}>0\right\}$ will happen,
a long position is taken in $\Delta _{k\Delta t}^{(\aleph,M)}$ forward contracts for some $N_{k\Delta t}^{(\aleph,M)}>0$ with maturity $(k+1)\Delta t$.
If at $k\Delta t$, $\aleph$ believes that $\left\{\xi_{n,k+1}^{(M)}<0\right\}$ will happen, a short position is taken in
$\Delta_{k\Delta t}^{(\aleph,M)}$ forward contracts with maturity $(k+1)\Delta t$.
Conditionally on $\mathcal{F}_k$, $\aleph$'s payoff\footnote{
When  $h(x)=g(x)=0$, this forward contract was introduced in \cite{Hu2020}.} at $(k+1)\Delta t$ is 
\begin{equation} 
P_{k\Delta t \rightarrow (k+1)\Delta t}^{(\aleph;forward)}= \Delta _{k\Delta t}^{(\aleph,M)}
\begin{cases}
        ( S_{(k+1)\Delta t}^{(n,h,g,u)}-S_{k\Delta t}^{(n,h,g)}e^{r_{k\Delta t}\Delta t} ), &\textup{w.p.} \; \mathbbm{p}^{(M)}_{n,\Delta t}p^{(\aleph,M)}_{\Delta t},  \\
        ( S_{k\Delta t}^{(n,h,g)}e^{r_{k\Delta t}\Delta t}-S_{(k+1)\Delta t}^{(n,h,g,d)} ), &\textup{w.p.} \; (1-\mathbbm{p}^{(M)}_{n,\Delta t})p^{(\aleph,M)}_{\Delta t},  \\
        ( S_{k\Delta t}^{(n,h,g)}e^{r_{k\Delta t}\Delta t}-S_{(k+1)\Delta t}^{(n,h,g,u)} ), &\textup{w.p.} \; \mathbbm{p}^{(M)}_{n,\Delta t}(1-p^{(\aleph,M)}_{\Delta t}),   \\
        ( S_{(k+1)\Delta t}^{(n,h,g,d)}-S_{k\Delta t}^{(n,h,g)}e^{r_{k\Delta t}\Delta t} ), &\textup{w.p.} \; (1-\mathbbm{p}^{(M)}_{n,\Delta t})(1-p^{(\aleph,M)}_{\Delta t}). 
\end{cases}
\label{eq30_p_forward}
\end{equation}
The conditional mean and variance of $P_{k\Delta t \rightarrow (k+1)\Delta t}^{(\aleph;forward)}$ are given by
\begin{align}
	\mathbb{E}\left(P_{k\Delta t \rightarrow (k+1)\Delta t}^{(\aleph;forward)}|\mathcal{F}_k\right)
	&= N_{k\Delta t}^{(\aleph,M)}(2p_{k\Delta t}^{(\aleph,M)}-1)\eta_{k,\Delta t}^{(M)}
		\left(\theta_{k,\Delta t}\Delta t+2 \sqrt{\mathbbm{p}_{n,\Delta t}^{(M)}(1-\mathbbm{p}_{n,\Delta t}^{(M)})\Delta t}\right),
	\label{eq31_meanandvar}
	\\
	\textup{Var}\left(P_{k\Delta t \rightarrow (k+1)\Delta t}^{(\aleph;forward)}|\mathcal{F}_k\right)
	&=  N_{k\Delta t}^{(\aleph,M)^2}{\eta_{k,\Delta t}^{(M)}}^2
		\left(1-4N_{k\Delta t}^{(\aleph,M)^2}(2p_{k\Delta t}^{(\aleph,M)}-1)^2\eta_{k,\Delta t}^2
		\mathbbm{p}_{n,\Delta t}^{(M)}(1-\mathbbm{p}_{n,\Delta t}^{(M)})\right)\Delta t,
	\nonumber
\end{align} 
where $\theta_{k\Delta t} =( \nu_{k\Delta t} - \frac{1}{2}{\eta_{k,\Delta t}^{(M)}}^2 - r_{k\Delta t} ) / \eta_{k,\Delta t}^{(M)} > 0$ can be interpreted
as the market price of risk in the binomial price process (\ref{eq29_S}) and $\eta_{k,\Delta t}^{(M)}$ is given by (\ref{eq22_eta}).
To guarantee that $\mathbb{E}\left(P_{k\Delta t \rightarrow (k+1)\Delta t}^{(\aleph;forward)}|\mathcal{F}_k\right) = O(\Delta t)$,
we set $p^{(\aleph,M)}_{k\Delta t} = (1+\lambda^{(\aleph,M)}\sqrt{\Delta t})/2$, for some $\lambda^{(\aleph,M)}>0$,
which we refer to as {\it $\aleph$'s information intensity}.\footnote{
The case of a misinformed trader can be considered in a similar manner.
	A misinformed trader with $p_{k\Delta t}^{(\aleph,M) }= (1-\lambda^{(\aleph,M)}\sqrt{\Delta t}) / 2$ trades long-forward (resp. short-forward)
	when the informed trader with $\lambda^{(\aleph,M)} > 0$ trades short-forward (resp. long-forward).
	A noisy trader (the trader with $p_{k\Delta t}^{(\aleph,M)} = 1/2$) will not trade any forward contracts, as he has no information about stock price direction.}
From (\ref{eq31_meanandvar}),
\begin{align*}
	\mathbb{E}\left(P_{k\Delta t \rightarrow (k+1)\Delta t}^{(\aleph;forward)}|\mathcal{F}_k\right)
	 	&= 2N_{k\Delta t}^{(\aleph,M)}\lambda^{(\aleph,M)}\eta_{k,\Delta t}^{(M)}\sqrt{\mathbbm{p}_{n,\Delta t}^{(M)}(1-\mathbbm{p}_{n,\Delta t}^{(M)})}\Delta t,
	 \nonumber
	\\
	\textup{Var}\left(P_{k\Delta t \rightarrow (k+1)\Delta t}^{(\aleph;forward)}|\mathcal{F}_k\right)
		&= N_{k\Delta t}^{(\aleph,M)^2}{\eta_{k,\Delta t}^{(M)}}^2\Delta t.
\end{align*}
The \textit{instantaneous information ratio} is given by
\begin{equation*}
	\textup{IR}\left(P_{k\Delta t \rightarrow (k+1)\Delta t}^{(\aleph,M;forward)}|\mathcal{F}_k\right)
	  = \frac{\mathbb{E}\left(P_{k\Delta t \rightarrow (k+1)\Delta t}^{(\aleph;forward)}|\mathcal{F}_k\right)}
		{\sqrt{\Delta t}\sqrt{\textup{Var}\left(P_{k\Delta t \rightarrow (k+1)\Delta t}^{(\aleph;forward)}|\mathcal{F}_k\right)}}
	 = 2\lambda^{(\aleph,M)}\sqrt{\mathbbm{p}_{n,\Delta t}^{(M)}(1-\mathbbm{p}_{n,\Delta t}^{(M)})}. 
\end{equation*}
Thus, the risk-adjusted payoff of $\aleph$’s forward strategy increases with the increase of $\aleph$’s information intensity $\lambda^{(\aleph,M)}$, as well as when $\mathbbm{p}_{n,\Delta t}^{(M)}$ approaches $1/2$.

\subsection{Option pricing under statistical arbitrage based on forward contracts}
\label{sec62}

\noindent Suppose $\aleph$ takes a short position in the option contract in the Black-Scholes-Merton market $\left(\mathcal{S},\mathcal{B},\mathcal{C}\right)$.\footnote{
The long position in the option contract is taken by a trader who trades the stock S with market perceived stock dynamics given by (\ref{eq18_Sconti}).}
The stock $\mathcal{S}$ has price dynamics  $S_t^{(h,g)},\;t\in[0,T]$, given by (\ref{eq18_Sconti});
the bond $\mathcal{B}$ has price dynamics $\beta_t,\;t\geq 0$ given by (\ref{eq28_bond});
and the option contract $\mathcal{C}$ has the price process $f_t=f(S_t,t),\;t\in [0,T]$ with terminal payoff $f_T=G(S_T)$.
When $\aleph$ trades the stock $\mathcal{S}$ to hedge the short position in $\mathcal{C}$, $\aleph$ simultaneously runs a forward contract strategy (Section \ref{sec61}).
This trading strategy (a combination of trading the stock and forward contracts) leads to an enhanced price process, with dynamics that can be expressed as follows:
conditionally on $\mathcal{F}_k$,
\begin{equation} 
S_{k+1,n}^{(h,g)}= 
\begin{cases}
	  S_{(k+1)\Delta t}^{(n,h,g,u)}+N_{k\Delta t}^{(\aleph,M)}  \left(S_{(k+1)\Delta t}^{(n,h,g,u)}-S_{k\Delta t}^{(n,h,g)}e^{r_{k\Delta t}\Delta t}\right),&\textup{w.p.} \; \mathbbm{p}^{(M)}_{n,\Delta t}p^{(\aleph,M)}_{\Delta t}, \\
      S_{(k+1)\Delta t}^{(n,h,g,d)}+N_{k\Delta t}^{(\aleph,M)}  \left(S_{k\Delta t}^{(n,h,g)}e^{r_{k\Delta t}\Delta t}-S_{(k+1)\Delta t}^{(n,h,g,d)}\right),&\textup{w.p.} \; (1-\mathbbm{p}^{(M)}_{n,\Delta t})p^{(\aleph,M)}_{\Delta t}, 
      \\
      S_{(k+1)\Delta t}^{(n,h,g,u)}+N_{k\Delta t}^{(\aleph,M)}  \left(S_{k\Delta t}^{(n,h,g)}e^{r_{k\Delta t}\Delta t}-S_{(k+1)\Delta t}^{(n,h,g,u)}\right),&\textup{w.p.} \; \mathbbm{p}^{(M)}_{n,\Delta t}(1-p^{(\aleph,M)}_{\Delta t}), 
      \\
      S_{(k+1)\Delta t}^{(n,h,g,d)}+N_{k\Delta t}^{(\aleph,M)}  \left((S_{(k+1)\Delta t}^{(n,h,g,d)}-S_{k\Delta t}^{(n,h,g)}e^{r_{k\Delta t}\Delta t}\right),&\textup{w.p.} \; (1-\mathbbm{p}^{(M)}_{n,\Delta t})(1-p^{(\aleph,M)}_{\Delta t}). 
\end{cases}
\label{eq32_s_forward}
\end{equation}
where $k =0,\ldots,n-1$, $n\Delta t = T$, and $S_{0,n}^{(h,g)}=S_0$.\footnote{
With every single share of the traded stock with price $S_{k\Delta t}^{(n,h,g)}$ at $k\Delta t$, $\aleph$
simultaneously enters $N_{k\Delta t}^{(\aleph,M)}$ forward contracts.
The forward contracts are long or short, depending on $\aleph$’s views on the sign of $\xi_{n,k+1}^{(M)}$.
It costs nothing to enter a forward contract at $k\Delta t$ with terminal time $(k+1)\Delta t$.}
Set $R_{k,n}^{(h,g)} = \ln\left(S_{k+1,n}^{(h,g)}/S_{k,n}^{(h,g)}\right)$ to be the stock log-return in period $[k\Delta t,(k+1)\Delta t)$.
Now its conditional mean and variance are 
\begin{align*}
\mathbb{E}\left(R_{k,n}^{(h,g)}|S_{k,n}^{(h,g)}\right) &= 
\left(\nu_{k \Delta t}+\frac{1}{2}{\eta_{k,\Delta t}^{(M)}}^2\right)\Delta t
-N_{k\Delta t}^{(\aleph,M)}\left(2p_{k\Delta t}^{(\aleph,M)}-1\right)
2\eta_{k,\Delta t}^{(M)}\sqrt{\mathbbm{p}_{n,\Delta t}^{(M)}(1-\mathbbm{p}_{n,\Delta t}^{(M)}\Delta t)}
\\
&+N_{k\Delta t}^{(\aleph,M)}\left(2p_{k\Delta t}^{(\aleph,M)}-1\right)
\left(\nu_{k \Delta t}-\frac{1}{2}{\eta_{k,\Delta t}^{(M)}}^2 - r_{k \Delta t}\right)\left(2\mathbbm{p}_{n,\Delta t}^{(M)}-1\right)\Delta t,
\\
\textup{Var}\left(R_{k,n}^{(h,g)}|S_{k,n}^{(h,g)}\right)&= {\eta_{k,\Delta t}^{(M)}}^2\Delta t 
\left(1+ N_{k\Delta t}^{(\aleph,M)}  \right)^2\left((1-\mathbbm{p}^{(M)}_{n,\Delta t})p^{(\aleph,M)}_{\Delta t}+ \mathbbm{p}^{(M)}_{n,\Delta t}(1-p^{(\aleph,M)}_{\Delta t}) \right)
\\
&+{\eta_{k,\Delta t}^{(M)}}^2\Delta t 
\left(1- N_{k\Delta t}^{(\aleph,M)}  \right)^2\left(\mathbbm{p}^{(M)}_{n,\Delta t}p^{(\aleph,M)}_{\Delta t}+ (1-\mathbbm{p}^{(M)}_{n,\Delta t})(1-p^{(\aleph,M)}_{\Delta t})\right)
\\
&-4{\eta_{k,\Delta t}^{(M)}}^2\Delta t 
N_{k\Delta t}^{(\aleph,M)^2}\left(2p_{k\Delta t}^{(\aleph,M)}-1 \right)^2{\eta_{k,\Delta t}^{(M)}}^2\mathbbm{p}_{n,\Delta t}^{(M)}(1-\mathbbm{p}_{n,\Delta t}^{(M)}).
\end{align*}
As in Section \ref{sec61}, we set $p^{(\aleph,M)}_{k\Delta t} = (1+\lambda^{(\aleph,M)}\sqrt{\Delta t})/2$ with $\lambda^{(\aleph,M)}>0$. Then
\begin{align*}
\mathbb{E}\left(R_{k,n}^{(h,g)}|S_{k,n}^{(h,g)}\right) &= 
\left(\nu_{k \Delta t}+\frac{1}{2}{\eta_{k,\Delta t}^{(M)}}^2+2N_{k\Delta t}^{(\aleph,M)}\lambda^{(\aleph,M)}\eta_{k,\Delta t}^{(M)}\sqrt{\mathbbm{p}_{n,\Delta t}^{(M)}(1-\mathbbm{p}_{n,\Delta t}^{(M)})}
\right)\Delta t,
\\
\textup{Var}\left(R_{k,n}^{(h,g)}|S_{k,n}^{(h,g)}\right)&= {\eta_{k,\Delta t}^{(M)}}^2\left(1+ {N_{k\Delta t}^{(\aleph,M)}}^2  \right) \Delta t.
\end{align*}
The instantaneous conditional market price of  risk is given by 
\begin{equation*}
\Theta\left(R_{k,n}^{(h,g)}|S_{k,n}^{(h,g)}\right) = \frac{\mathbb{E}\left(R_{k,n}^{(h,g)}|S_{k,n}^{(h,g)}\right)-r_{k \Delta t}\Delta t}{\sqrt{\Delta t}\textup{Var}\left(R_{k,n}^{(h,g)}|S_{k,n}^{(h,g)}\right)} = \frac{\theta_{k\Delta t}+2N_{k\Delta t}^{(\aleph,M)}\lambda^{(\aleph,M)}\sqrt{\mathbbm{p}_{n,\Delta t}^{(M)}(1-\mathbbm{p}_{n,\Delta t}^{(M)})}}{\sqrt{1+N_{k\Delta t}^{(\aleph,M)^2}}},
\end{equation*}
where $\theta_{k\Delta t} =\left( \nu_{k\Delta t} - \frac{1}{2}{\eta_{k,\Delta t}^{(M)}}^2-r_{k\Delta t} \right) / \eta_{k,\Delta t}^{(M)}$.
The value
\begin{equation*}
	N_{k\Delta t}^{(\aleph,M;\text{opt})} = 2\frac{\lambda^{(\aleph,M)}}{\theta_{k\Delta t}}\sqrt{\mathbbm{p}_{n,\Delta t}^{(M)}(1-\mathbbm{p}_{n,\Delta t}^{(M)})},
\end{equation*}
produces the maximum (optimal) instantaneous market price of risk,
\begin{equation*}
\Theta^{(\text{opt})}\left(R_{k,n}^{(h,g)}|S_{k,n}^{(h,g)}\right)
	= \sqrt{{\theta_{k\Delta t}}^2+4{\lambda^{(\aleph,M)}}^2\mathbbm{p}_{n,\Delta t}^{(M)}(1-\mathbbm{p}_{n,\Delta t}^{(M)})}.
\end{equation*}
With $N_{k\Delta t}^{(\aleph,M)} = N_{k\Delta t}^{(\aleph,M;\text{opt})}$, the conditional mean  and variance have the simpler representations: 
\begin{align*}
\mathbb{E}\left(R_{k,n}^{(h,g)}|S_{k,n}^{(h,g)}\right)
	&= 	\left(\nu_{k \Delta t}+\frac{1}{2}{\eta_{k,\Delta t}^{(M)}}^2
	+ 4\frac{{\lambda^{(\aleph,M)}}^2}{\theta_{k\Delta t}}\eta_{k,\Delta t}^{(M)}\mathbbm{p}_{n,\Delta t}^{(M)}(1-\mathbbm{p}_{n,\Delta t}^{(M)})
	\right)\Delta t,
	\\
	\textup{Var}\left(R_{k,n}^{(h,g)}|S_{k,n}^{(h,g)}\right)
	&= {\eta_{k,\Delta t}^{(M)}}^2\left(1+ 4\frac{{\lambda^{(\aleph,M)}}^2}{\theta^2_{k\Delta t}} \mathbbm{p}_{n,\Delta t}^{(M)}(1-\mathbbm{p}_{n,\Delta t}^{(M)}) 			\right)\Delta t .
\end{align*}

$\aleph$'s dividend $D_{k\Delta t}^{(\aleph)}$  in $[k\Delta t,(k+1)\Delta t)$, generated by the enhanced price process (\ref{eq32_s_forward})
with $N_{k\Delta t}^{(\aleph,M)} = N_{k\Delta t}^{(\aleph,M;\text{opt})}$, is
\begin{equation*}
D_{k\Delta t}^{(\aleph)} = \eta_{k,\Delta t}^{(M)}\left(\sqrt{\theta_{k\Delta t}^2+4{\lambda^{(\aleph,M)}}^2
\mathbbm{p}_{n,\Delta t}^{(M)}(1-
\mathbbm{p}_{n,\Delta t}^{(M)})}-\theta_{k\Delta t}\right).
\end{equation*}
Expressed in terms of the dividend, the optimal instantaneous market price of risk is\footnote{
	If $\aleph$ is a misinformed trader, the opposite of an informed trader is followed, that is, what is a profit for the informed trader will be a loss for the misinformed trader.
	Thus, in general, if $ p_{\Delta t}^{(\aleph,M)} = (1+\lambda^{(\aleph,M)}\sqrt{\Delta t})/2$, for some $\lambda^{(\aleph,M)}\in R$,
	the dividend yield $D_{k\Delta t}^{(\aleph)}\in R$, is given by
	$D_{k\Delta t}^{(\aleph)} = \textup{sign}\left(\lambda^{(\aleph,M)}\right)\eta_{k,\Delta t}^{(M)}\left(\sqrt{\theta_{k\Delta t}^2+4{\lambda^{(\aleph,M)}}^2
	\mathbbm{p}_{n,\Delta t}^{(M)}(1-\mathbbm{p}_{n,\Delta t}^{(M)})}-\theta_{k\Delta t}\right)$,
	where $\textup{sign}\left(\lambda^{(\aleph,M)}\right) =1$, if $\lambda^{(\aleph,M)}>0$; $\textup{sign}\left(\lambda^{(\aleph,M)}\right) =0$, if $\lambda^{(\aleph,M)}=0$;
	and $\textup{sign}\left(\lambda^{(\aleph,M)}\right) =-1$, if $\lambda^{(\aleph,M)}<0$.
}
\begin{equation*}
\Theta^{(\text{opt})}\left(R_{k,n}^{(h,g)}|S_{k,n}^{(h,g)}\right)=\left( \nu_{k\Delta t}+D_{k\Delta t}^{(\aleph)} - \frac{1}{2}{\eta_{k,\Delta t}^{(M)}}^2-r_{k\Delta t} \right) / \eta_{k,\Delta t}^{(M)}.
\end{equation*}

$\aleph$’s mean rate of log-return (when $\aleph$ trades $\mathcal{S}$ using the enhanced price  process (\ref{eq32_s_forward}) with $N_{k\Delta t}^{(\aleph,M)}= N_{k\Delta t}^{(\aleph,M;\text{opt})}$) is
$\nu^{(\aleph)}_{k \Delta t} = \nu_{k \Delta t}+D_{k\Delta t}^{(\aleph)}$.
Then
\begin{equation}
\nu^{(\aleph)}_{k \Delta t} = \sqrt{\left(\nu_{k \Delta t}-\frac{1}{2} {\eta_{k,\Delta t}^{(M)}}^2- r_{k \Delta t}\right)^2+4{\lambda^{(\aleph,M)}}^2\mathbbm{p}_{n,\Delta t}^{(M)}(1-\mathbbm{p}_{n,\Delta t}^{(M)})  }+\frac{1}{2}{\eta_{k,\Delta t}^{(M)}}^2+r_{k \Delta t}.
\label{eq33_nu}
\end{equation}
Following the arguments in the derivation of (\ref{eq25_qp}),  we obtain the conditional risk-neutral probabilities 
\begin{equation}
	\mathbbm{q}_{k,\Delta t}^{(S,\aleph)}
	 = \mathbbm{p}_{n,\Delta t}^{(M)} - \theta^{(S,\aleph)}_{k,\Delta t}\sqrt{\mathbbm{p}_{n,\Delta t}^{(M)}(1-\mathbbm{p}_{n,\Delta t}^{(M)})\Delta t},\quad
	    \theta^{(S,\aleph)}_{k,\Delta t} = \frac{\nu^{(\aleph)}_{k \Delta t}-r_{k \Delta t}}{\eta^{(M)}_{k,\Delta t}},\;
	     k = 0, \ldots, n-1,\; n \in \mathcal{N}.
	\label{eq34_qp}
\end{equation}
Conditioned on $\mathcal{F}_k$, the stock risk-neutral dynamics is given by
\begin{equation}
	S_{(k+1)\Delta t}^{(n,\aleph,\mathbbm{q})}=
	\begin{cases}
	S_{(k+1)\Delta t}^{(n,\aleph,\mathbbm{q},u)}&=S_{k\Delta t}^{(n,\aleph,\mathbbm{q})}\exp\left\{\nu_{k \Delta t} \Delta t+\eta^{(M)}_{k,\Delta t}\sqrt{\Delta t}\sqrt{\frac{1-\mathbbm{p}^{(M)}_{n,\Delta t}}{\mathbbm{p}^{(M)}_{n,\Delta t}}} \right\},\;\textup{w.p.}\;\mathbbm{q}^{(S,\aleph)}_{k,\Delta t},
	\\
	S_{(k+1)\Delta t}^{(n,\aleph,\mathbbm{q},d)}&=S_{k\Delta t}^{(n,\aleph,\mathbbm{q})}\exp\left\{\nu_{k \Delta t} \Delta t-\eta^{(M)}_{k,\Delta t}\sqrt{\Delta t}\sqrt{\frac{\mathbbm{p}^{(M)}_{n,\Delta t}}{1-\mathbbm{p}^{(M)}_{n,\Delta t}}}\right\},\;\textup{w.p.}\;1-\mathbbm{q}^{(S,\aleph)}_{k,\Delta t},
	\end{cases}
	\label{eq35_Sq}
\end{equation}
for $k = 0, \dots , n-1$ with $S_0^{(n,\aleph,\mathbbm{q})} = S_0$.
According to (\ref{eq35_Sq}), $\aleph$ hedges only the risk for the stock movements.
$\aleph$ has chosen the best allocation $N_{k\Delta t}^{(\aleph,M)} = N_{k\Delta t}^{(\aleph,M;\text{opt})}$ in the enhanced price process (\ref{eq32_s_forward}) when the stock is traded.
$\aleph$ leaves the  risk of wrong bets unhedged.
In this way,  $\aleph$'s forward strategy does not lead to a pure arbitrage opportunity for $\aleph$.
Rather $\aleph$’s strategy (based on his information about the stock-price movement)  leads to “statistical arbitrage”, as those stocks with improved market price for risk $\Theta^{(\text{opt})}\left(R_{k,n}^{(h,g)}|S_{k,n}^{(h,g)}\right)>\theta_{k\Delta t}$ are traded.
The risk-neutral price $f_{k \Delta t}^{(n,\aleph)} =f\left( S_{k \Delta t}^{(n,\aleph,\mathbbm{q})},k\Delta t\right)$ of the option $\mathcal{C}$ is given by
\begin{equation}
	f_{k \Delta t}^{(n,\aleph)} = e^{-r\Delta t}\left\{\mathbbm{q}^{(S,\aleph)}_{k,\Delta t}f_{(k+1) \Delta t}^{(n,\aleph,u)}
							+(1-\mathbbm{q}^{(S,\aleph)}_{k,\Delta t})f_{(k+1) \Delta t}^{(n,\aleph,d)}\right\},
	\label{eq39_ft}
\end{equation}
where $f_{(k+1) \Delta t}^{(n,\aleph,u)} =f\left( S_{(k+1) \Delta t}^{(n,\aleph,\mathbbm{q},u)},(k+1)\Delta t\right)$, $f_{(k+1) \Delta t}^{(n,\aleph,d)}
=f\left( S_{(k+1) \Delta t}^{(n,\aleph,\mathbbm{q},d)},(k+1)\Delta t\right)$,
$k = 0,\ldots\ldots,n-1$, and $f_{n \Delta t}^{(n,\aleph)} = f_{T}^{(n,\aleph)} = G(S_T)$.

\subsection{Implied information intensity  $\mathbf{\lambda^{(\aleph,M)}}$}
\label{sec63}
\begin{figure}
 	\begin{center} 
    	\centering\includegraphics[width=0.5\textwidth]{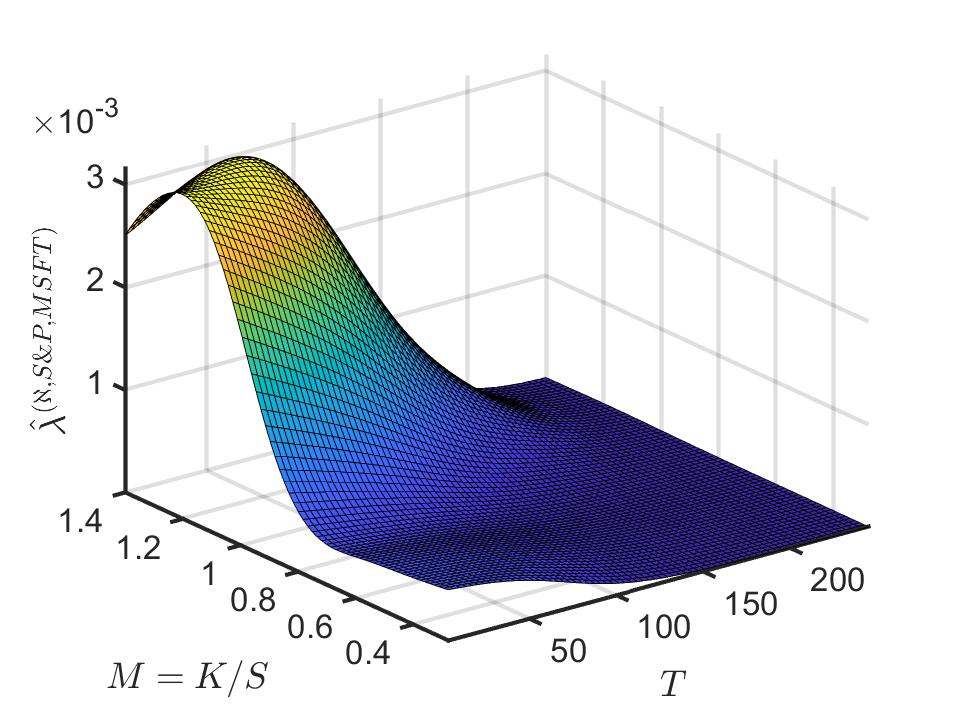}
    	\caption{Implied information intensity against time to maturity $T$ and moneyness $M=K/S$.}
    	\label{fig10_lambda}
    \end{center}
\end{figure}

\noindent We use numerical data to illustrate the method to find the implied information intensity $\lambda^{(\aleph,M)}>0$ defined in Section \ref{sec61}.
As in previous numerical examples in this paper, we consider $\lambda^{(\aleph,M)}=\lambda^{(\aleph,\text{S\&P,MSFT})}$,
the implied information rate of a MSFT option trader using market information from the S\&P 500 index.
Following the framework in Section \ref{sec62} for an informed trader, we construct a binomial option pricing tree using the data\footnote{
	The data set includes MSFT price data over the period $07/01/2018$ to $06/30/2020$ (Yahoo Finance);
	S\&P 500 price data from $07/01/2019$ to $06/30/2020$ (Yahoo Finance);
	10-year Treasury rate from $07/01/2019$ to $06/30/2020$ (U.S. Department of the Treasury);
	and the call option price data for the MSFT at $07/01/2019$ (CBOE).}
from Section \ref{sec5} and the estimated results in Table \ref{tab1_est_para}.
Then in (\ref{eq33_nu})-(\ref{eq39_ft}): $\eta_{k,\Delta t}^{(M)} = \eta_{k,\Delta t}^{(\text{S\&P})}$ with $\mathbbm{p}^{(M)}_{n,\Delta t} = \mathbbm{p}^{(\text{S\&P})}_{n,\Delta t}$
as in (\ref{eq_eta53}); $\nu_{k\Delta t}$ are estimated using the mean log-return for a one-year window of MSFT price data ending on $k\Delta t$;
and $r_{k\Delta t}$ is the risk-free rate at time $k\Delta t$.
For the informed trader's call option placed on 07/01/2019, we construct the option price tree
$C^{(\text{MSFT},\aleph)}_t(S^{(\text{MSFT})}_{0},K,T-t,\hat{\nu},\hat{\sigma},\hat{\gamma},\hat{\delta},\lambda^{(\aleph,\text{S\&P,MSFT})})$.
Here, the terminal time $T$ is 252 trading days (06/30/2020) and $t$ varies from 0 (07/01/2019) through 252.
Denoting CBOE's call option price data of MSFT on $07/01/2019$ as $C^{(\text{MSFT})} = C^{(\text{MSFT})}_t(K,T-t)$, for different strike price $K$ and $t$,  we calculate
\begin{equation*}
	\hat{\lambda}^{(\aleph,\text{S\&P,MSFT})} = \textup{arg\;min}_{\lambda^{(\aleph,\text{S\&P,MSFT})}>0} \left\{\left(\frac{C^{(\text{MSFT},\aleph)}-C^{(\text{MSFT})}}{C^{(\text{MSFT})}}\right)^2\right\}.
\end{equation*}
	Figure \ref{fig10_lambda} exhibits the implied $\hat{\lambda}^{(\aleph,\text{S\&P,MSFT})}$ surface against time to maturity and moneyness.
	The values of $\hat{\lambda}^{(\aleph,\text{S\&P,MSFT})}$ vary from $2.25 \times 10^{-6}$ to $3.2 \times 10^{-3}$, for $M\in[0.28,1.4]$.
The value $M = 1.12$ represents a peak in $\hat{\lambda}^{ (\aleph,M) }$. $\hat{\lambda}^{ (\aleph,M) }$ drops steeply from this peak value for the first five months,
after which it slowly decreases.
In general, the value of $\hat{\lambda}^{(\aleph,\text{S\&P,MSFT})}$ when $M\in[1,1.4]$ is higher than that when $M\in[0.28,1]$. 

\section{Conclusion}
\label{sec7}

\noindent
Inclusion of more information on market microstructure will inevitably lead to better option pricing models.
As noted in the survey by \cite{Easley2003}, while ``market microstructure and asset pricing models both consider the behavior and formation
of prices in asset markets ... neither literature explicitly recognizes the importance and role of the factors so crucial to the other approach."
As an example of the separation in these two approaches, they contrast the conclusion that `reasonable stationarity assumptions in asset pricing
theory lead to the expectation that price changes follow a random walk' with a quote from \cite{Hasbrouk1996}
	 ``At the level of transaction prices, ..., the random walk conjecture is a straw man, a hypothesis that is very easy to reject in most markets even
	in small data samples.
	In microstructure, the question is not ``whether'' transaction prices diverge from a random walk, but rather ``how much'' and ``why?''.
In this paper, we have taken two steps to increase the amount of microstructure information included in a discrete, binomial option pricing model.

Using the Donsker-Prokhorov invariance principle, we extended the KSRF model to allow for
variably-spaced trading instances, which, for example, is of critical importance to the short seller of an option.
In particular we derive the expression for the discrete risk-neutral upturn probability $q_{n,k}$ for an arbitrarily spaced time period
and quantify it's dependence on the discrete, time-varying, natural-world upturn probability $p_{n,k}$.
By reverting to a fixed-spaced trading interval (of 1 day), and using market data, we compare the relative behaviors of $q_{n,k}$ and $p_{n,k}$.
We have also inferred the underlying natural probabilities $p_{\Delta t}^{(\text{CRR})}$ and $p_{\Delta t}^{(\text{JR})}$ that are missing from the
Cox-Ross-Rubenstein and Jarrow-Rudd models and compare them with actual probabilities.

We employed the Cherny-Shiryaev-Yor invariance principle to construct option pricing within a complete market model in which the underlying stock
dynamics depends on the (path-dependent) history of a market index (or market influencing factor).
We add terms to the discrete log-return dynamics which, in the continuum limit, correspond to path-integrated volatility and a doubly-integrated volatility.
A critical variable that emerges from this extension is $\xi_{n,k}^{(M)}$ defined in (\ref{eq13_setting_CSYip}(b)),
which depends on the upturn probability $\mathbbm{p}_{\Delta t}$ of the underlying stock.
The dynamics of $\xi_{n,k}^{(M)}$ and the added terms are investigated numerically in the Appendix.
As a result of the path dependence, the market price of risk, which drives the risk-neutral probability, is time-dependent,
leading to a heavy-tailed distribution for the log-returns and potentially a more realistic pricing model for an option.
Using numerical data, we explore the potential deviation between the price volatility of a call option based upon this new asset pricing
approach and that provided by the CBOE.
With reference to the first paragraph of this section, we note that, in our model based upon the Cherny-Shiryaev-Yor invariance principle, the
discrete time dynamics of the stock prices does not follow a random walk, as the stock price is path dependent.

We explored the implications for option pricing under a market in which traders employ a
statistical arbitrage strategy using forward contracts based upon their assumed knowledge of the market index upturn probability.
In such a market, the market price of risk develops a drift term that now reflects the information intensity $\lambda^{(\aleph,M)}$
of the traders.
Using a numerical example, we explore the value of the information intensity of a call option as a function of moneyness and time to
maturity.
This example explicitly demonstrates that the market microstructure view -- `in real markets, traders use information about stock price
direction' -- can, in fact, be incorporated within dynamic asset pricing theory.
Thus we believe that reconciling market microstructure and dynamic asset pricing is possible, and that one approach to doing so is through
discrete-time pricing models.
In particular, an idea suggested by \cite{Sahalia2014} to model microstructure dynamics as discrete semimartingales plus noise provides one
plausible route.

\bibliography{Refs_Hu_et_al}
\bibliographystyle{apalike}

\appendix
\section{Appendix}

\noindent
We examine the discrete log-return process given by (\ref{eq19_Rnhgs}) in the context of the numerical example of Section \ref{sec44}.
The total return of a stock through time $k\Delta t$ is modeled as
\begin{equation}
	R_{k\Delta t}^{(n,h,g,S)}  = \nu k \Delta t + \sigma X^{(n)}_{k/n} + \gamma Y^{(n,h)}_{k/n} + \delta V^{(n,g)}_{k/n},
	\label{disc_rtn}
\end{equation}
where the terms on the right-hand side are computed based upon the dynamics of the S\&P 500 index.
We consider the values of the S\&P 500 index from 07/01/1986 through 06/30/2020 and compute log-returns of the index to evaluate
the daily time series $\xi_{n,k}^{(\text{S\&P})}, X^{(n)}_{k/n}$, $Y^{(n,h)}_{k/n}$ and $ V^{(n,g)}_{k/n}$ from equations
(\ref{eq13_setting_CSYip}(b-e)).
We consider the predictability of the model (\ref{disc_rtn}) over a 1-year interval ($k \in [1,252]$).
With 33 years of S\&P data, we considered 30 non-overlapping (and thus somewhat independent) 1-year periods of S\&P samples.
In choosing the coefficients $\nu, \sigma, \gamma, \sigma_h, \delta$ and $\sigma_g$ needed to evaluate the terms in (\ref{disc_rtn}),
we used fitted values from row S\&P${}^a$ of Table \ref{tab1_est_para}.

\begin{figure}[ht]
    \centering
    \subcaptionbox{$X^{(n)}_{k/n}$ }{\includegraphics[width=0.32\textwidth]{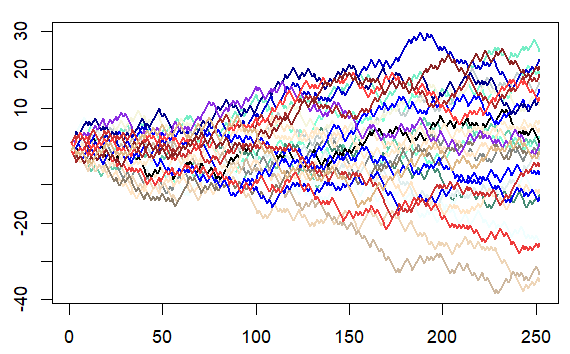}}\hspace{0em}%
    \subcaptionbox{$Y^{(n,h)}_{k/n}$}{\includegraphics[width=0.32\textwidth]{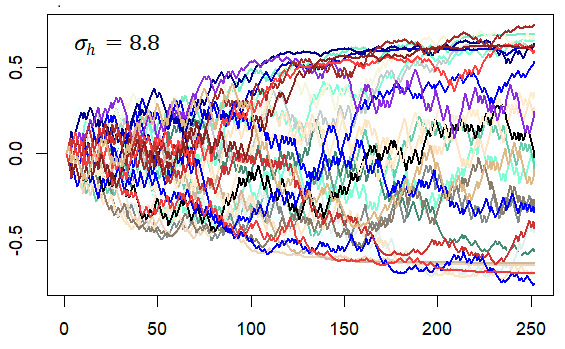}}\hspace{0em}%
    \subcaptionbox{$V^{(n,g)}_{k/n}$}{\includegraphics[width=0.32\textwidth]{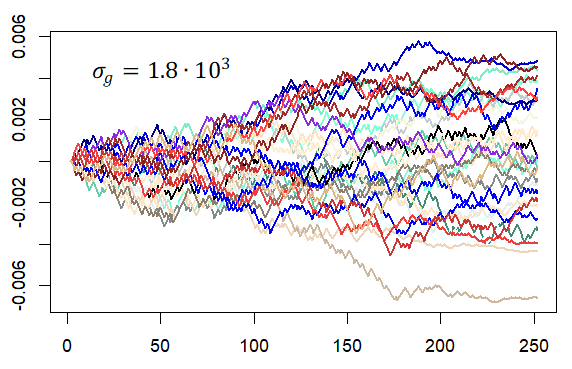}}\hspace{0em}%
    \caption{One-year times series traces of $ X^{(n)}_{k/n},  Y^{(n,h)}_{k/n}, V^{(n,g)}_{k/n}$ and $S^{(n,h,g,S)}_{k \Delta t}$ computed from the S\&P 500 data.}
    \label{fig_XYV}
\end{figure}

\begin{figure}[ht]
    \centering
    \subcaptionbox{$\sigma_g = 1$}{\includegraphics[width=0.24\textwidth]{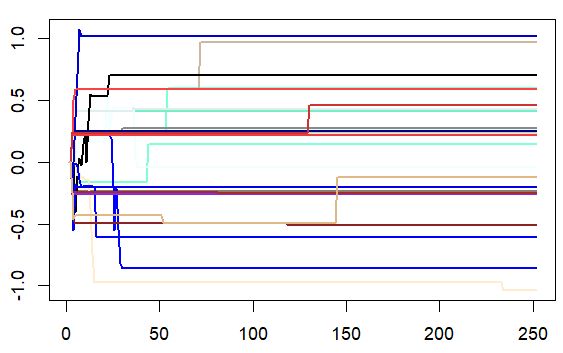}}\hspace{0em}%
    \subcaptionbox{$\sigma_g = 10$}{\includegraphics[width=0.24\textwidth]{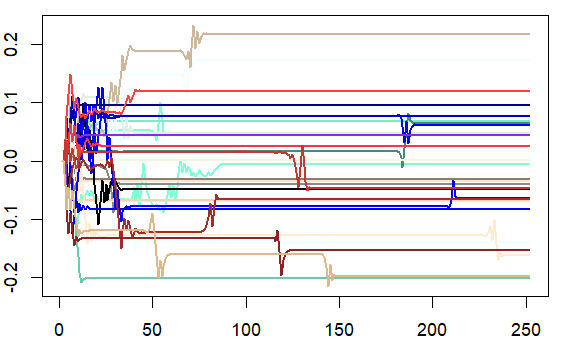}}\hspace{0em}%
    \subcaptionbox{$\sigma_g = 10^2$}{\includegraphics[width=0.244\textwidth]{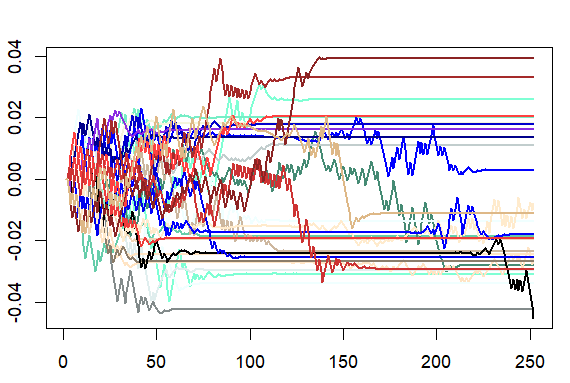}}\hspace{0em}%
    \subcaptionbox{$\sigma_g = 10^3$}{\includegraphics[width=0.242\textwidth]{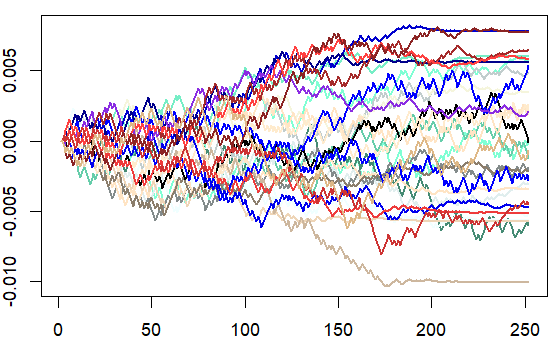}}\hspace{0em}%
    \caption{Computation of $V^{(n,g)}_{k/n}$ with various values of $\sigma_g$ showing the band-pass filter effect of the Gaussian function $g()$ employed here.}
    \label{fig_sgmg}
\end{figure}

Figure \ref{fig_XYV} displays the time-series samples for the terms $X^{(n)}_{k/n}$, $Y^{(n,h)}_{k/n}$ and $V^{(n,g)}_{k/n}$ in (\ref{disc_rtn}).
$ X^{(n)}_{k/n}$ has the characteristics of a BM (to which it should weakly converge in the limit $\Delta t \downarrow 0$).
As $\Delta t = 1$ for this numerical data, the $\sqrt t$ growth of the BM becomes significant. (See the vertical axis in Figure \ref{fig_XYV}(a).)
Thus fitted values of $\sigma$ must be $< O(10^{-2})$ to “control” this growth.
Given the similar growth with time in the magnitude of the arguments of $h()$ and $g()$ (not shown),
it is clear that the Gaussian form chosen here for $h()$ and $g()$  acts as a band-pass filter,
selectively passing argument values within one to two $\sigma_h$ (respectively $\sigma_g$) of zero and exponentially suppressing higher magnitude arguments.
To demonstrate this filtering effect, $V^{(n,g)}_{k/n}$ is evaluated with $\sigma_g$ taking on values $1, 10, 10^2$ and $10^3$.
The results are shown in Figure \ref{fig_sgmg}.
Thus it would seem critical to appropriately capture the variance of these arguments in chosing the functions $h()$ and $g()$.
The fitted values $\sigma_h = 8.8$ and $\sigma_g = 1.8 \cdot 10^3$ obtained in Table \ref{tab1_est_para} suggest that $h()$ and $g()$ are most appropriately
chosen to act as very wide band-pass filters for the S\&P data.

\begin{figure}[htb]
    \centering
    \subcaptionbox{$S^{(n,h,g,S)}_{k \Delta t}$}{\includegraphics[width=0.45\textwidth]{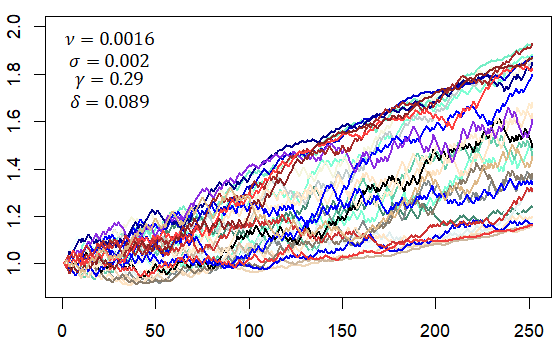}}\hspace{0em}%
    \subcaptionbox{MSFT}{\includegraphics[width=0.46\textwidth]{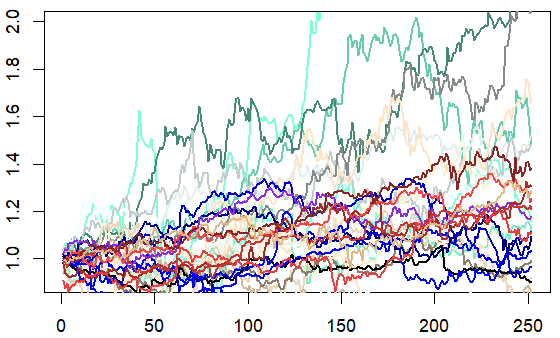}}\hspace{0em}%
    \caption{One-year times series traces of $S^{(n,h,g,S)}_{k \Delta t}$ computed from (\ref{disc_rtn}) based upon S\&P 500 index data
compared to actual price traces of MSFT for the same 1-year periods.}
    \label{fig_Sk}
\end{figure}
Figure \ref{fig_Sk} displays samples of 1-year traces for the cumulative stock price $S^{(n,h,g,S)}_{k \Delta t}$ obtained from the
S\&P-based model (\ref{disc_rtn}) and compares them with actual cumulative price traces of MSFT computed for the same 1-year periods.
(The computations assumed an initial starting price of $S_0 = 1$ for each trace.)
Traces for the same time period are colored the same in Figures \ref{fig_Sk}(a) and (b).
 
As noted in Section \ref{sec4}, as $\Delta t \downarrow 0$, the processes (\ref{eq14_BC}) converge weakly to, respectively,
a BM $\mathbb{B}_{[0,T]}=\left\{B_t\right\}_{0\leq t\leq T}$ on $[0,T]$
and to $\mathbb{C}_{[0,T]}^{(h)}=\left\{C_t^{(h)}=\int_0^t h\left(B_s\right)d B_s\right\}_{0 \leq t\leq T}$;
the process $\mathbb{G}^{(n,g)}_{[0,T]}$ converges weakly to
$\left\{ G_{t}^{(g)} = \int_0^t g(\int_0^{\nu} B_{u}du)dB_{\nu} \right\}_{0 \le t \le T}$;
and the discrete stock price process (\ref{eq17_Snhgs}) converges weakly to (\ref{eq18_Sconti}).
We examine the behavior of $B_t, C_t^{(h)}, G_t^{(g)}$, and $S_t^{(n,h,g,S)}$ over the interval $t \in [0,1]$ by assuming
$h(x)$ and $g(x)$ have the same Gaussian forms as in the discrete model.

\begin{figure}[ht]
    \centering
    \subcaptionbox{$B_t$}{\includegraphics[width=0.32\textwidth]{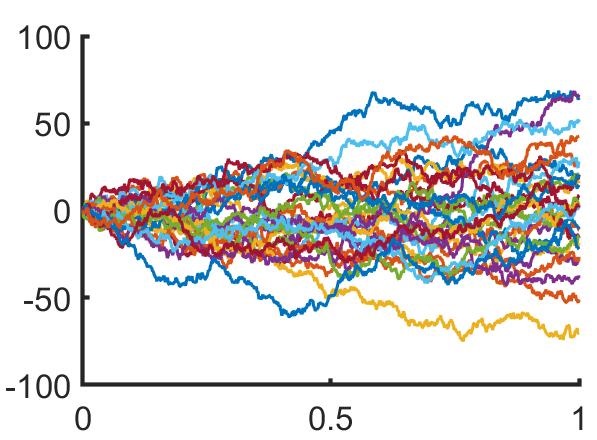}}\hspace{0em}%
    \subcaptionbox{$C_t$}{\includegraphics[width=0.31\textwidth]{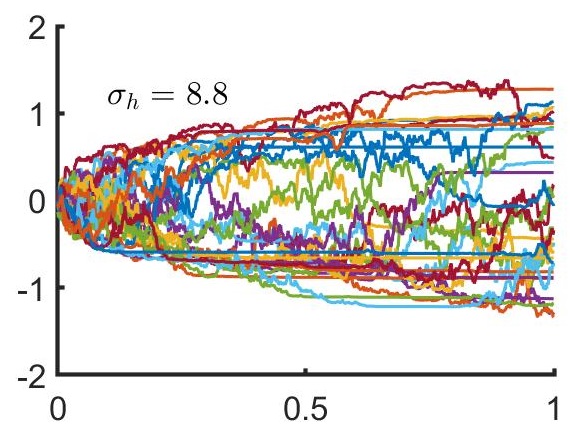}}\hspace{0em}%
    \subcaptionbox{$G_t$}{\includegraphics[width=0.31\textwidth]{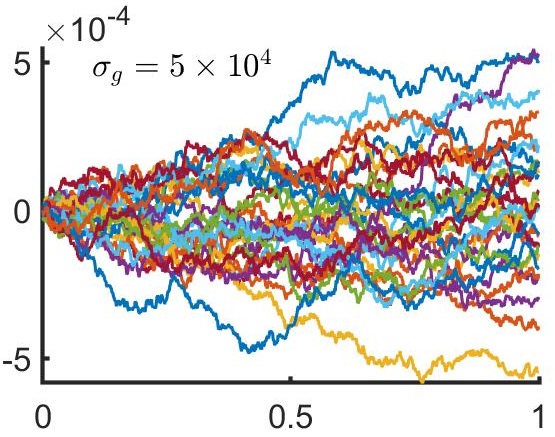}}\hspace{0em}%
    \subcaptionbox{$S_t$: $\gamma = 0.1, \delta = 0.1$}{\includegraphics[width=0.32\textwidth]{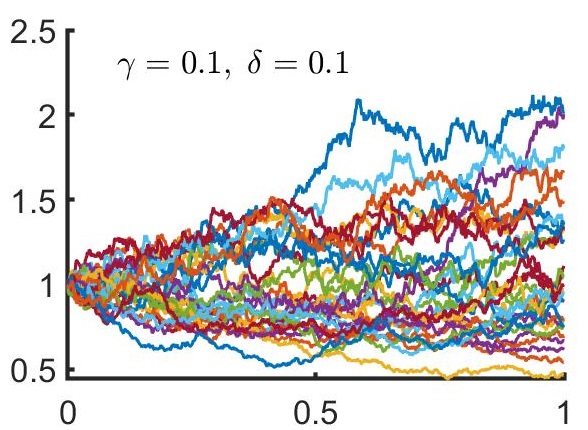}}\hspace{0em}%
    \subcaptionbox{$S_t$: $\gamma = 0.5, \delta = 0.1$}{\includegraphics[width=0.31\textwidth]{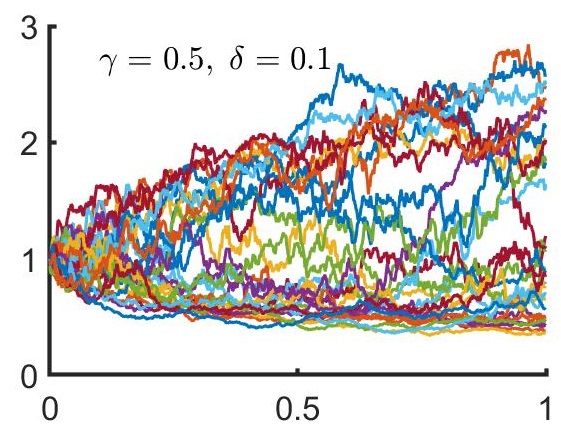}}\hspace{0em}%
    \subcaptionbox{$S_t$: $\gamma = 0.5, \delta = 10^3$}{\includegraphics[width=0.31\textwidth]{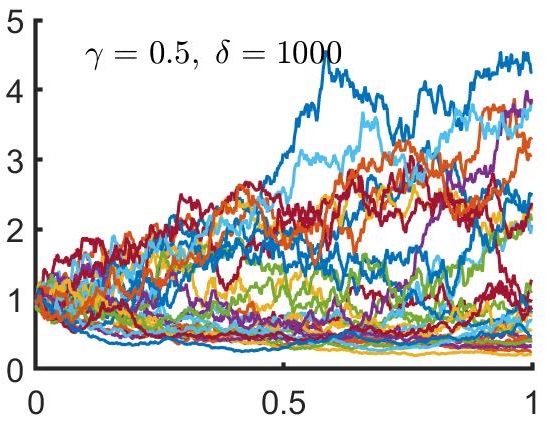}}\hspace{0em}%
    \caption{Times series traces of $ B_t,  C_t, G_t$ and $S^{(n,h,g,S)}_ t$ computed from the continuum model.}
    \label{fig_BCGS}
\end{figure}

The behavior of a BM is well known, Figure \ref{fig_BCGS}(a) displays 30 traces of a BM over the time period $[0,1]$ computed with $\Delta t = 10^{-3}$.
The behaviors of $C_t^{(h)}$ and $G_t^{(g)}$ computed from these trajectories are shown in Figure \ref{fig_BCGS}(b,c).
Value of $\sigma_h = 8.8$ and $\sigma_g = 5\cdot 10^4$ were used for $h()$ and $g()$.
Cumlative price series computed with $\nu = 0, \sigma = 0.01$ and indicated values for $\gamma$ and $\delta$ are showin in Figure \ref{fig_BCGS}(d-f).

\end{document}